\documentclass[manuscript]{acmart}

\AtBeginDocument{%
  }

\usepackage{float}
\usepackage{graphicx}
\usepackage{subcaption}
\usepackage{textcomp}
\usepackage{xcolor}
\usepackage{pifont}
\usepackage{tikz}

\newcommand*\filledcircled[1]{%
  \tikz[baseline=(char.base)]{%
    \node[
      circle,
      draw=black,        
      fill=black,        
      text=white,        
      font=\scriptsize,  
      inner sep=0.5pt,   
      minimum size=1em   
    ] (char) {#1};
  }%
}

\begin{document}

\title[Scaling and Placement for Serverless Streaming]{CLASP: Chained-Request-Aware Scaling and Operator Placement for Serverless Stream Processing}

\author{Tianyu Qi}
\email{tiqi@student.unimelb.edu.au}
\orcid{0009-0006-1616-9902}

\author{Maria A. Rodriguez}
\email{marodriguez@unimelb.edu.au}
\orcid{0000-0002-2831-8526}

\author{Rajkumar Buyya}
\email{rbuyya@unimelb.edu.au}
\orcid{0000-0001-9754-6496}
\affiliation{%
  \institution{Quantum Cloud Computing and Distributed Systems (qCLOUDS) Lab,
    School of Computing and Information Systems, The University of Melbourne}
  \city{Melbourne}
  \state{Victoria}
  \country{Australia}
}

\renewcommand{\shortauthors}{Qi et al.}
\acmArticleType{Research}



\begin{abstract}

Stateful serverless (Function-as-a-Service) environments, whose workers host state servers, are increasingly used for stream processing. A stream application is a pipeline of operators, where each operator forwards intermediate data downstream through a chained request. As input rates fluctuate, the system should adjust operator parallelism and place instances across workers to sustain the incoming rate. Existing approaches do so without fully accounting for chained-request overhead, leading them to misestimate the required number of workers. Too few leave the cluster unable to keep up with the input rate, while too many route a larger fraction of chained requests across worker boundaries, increasing end-to-end latency.

We propose CLASP, a scaling and scheduling strategy for stream processing in stateful serverless environments. At runtime, CLASP estimates execution cost and chained-request cost from observed metrics. Under a capacity model that covers the two costs, it adjusts operator parallelism and packs operators onto the fewest workers that can sustain the target input rate. Once a scaling decision is made, CLASP migrates each operator's state together with its instances, thereby minimizing execution pause time. Experiments show that CLASP improves throughput by up to 3.3$\times$ and reduces median end-to-end latency by up to 76\% compared with state-of-the-art scaling strategies.

\end{abstract}

\maketitle

\section{Introduction}

Serverless computing has emerged as a compelling cloud execution model, as it offloads infrastructure management to the provider and elastically scales lightweight execution units in response to workload fluctuations~\cite{cardellini22streamsurvey}. These properties align closely with the requirements of stream processing applications, which demand low-latency execution, elastic scalability, and minimal operational overhead~\cite{song23sponge}. Driven by this alignment, extensive research has investigated the deployment of streaming applications on serverless platforms, with representative use cases including video processing~\cite{ao2018serverlessvideo}, data analytics~\cite{nastic2017serverlessstream}, and IoT sensor monitoring~\cite{hall2019execution, cicconetti2020decentralized, yao2023performance}.

A stream processing application is modeled as a directed acyclic graph (DAG) of operators, where each operator processes incoming data and forwards results to its downstream operators~\cite{hirzel2014streamoptimizations}. In serverless environments, these operators are deployed as ephemeral functions, and inter-operator data transfer is realized through chained function invocations, where each function instance invokes one or more downstream functions upon completing its local processing~\cite{li2022faasflow}. We refer to such an invocation, together with the intermediate data it carries, as a chained request. To avoid the latency of forwarding chained requests through a centralized scheduler, distributed scheduling is commonly adopted in serverless DAG applications, where each worker maintains a local scheduler that independently dispatches its locally generated chained requests to the destination workers~\cite{wang2025octopus,zhang2025decentralizedschedulingweb,chen2025ekko}. Additionally, many operators are intrinsically stateful, relying on the state accumulated from prior events to process incoming data~\cite{xu2023statefulapplication}. Since function instances are ephemeral in serverless environments, operator state must be persisted in external storage. To reduce state access time, stateful serverless platforms are commonly used for stream processing, which retain the operator state on the worker nodes~\cite{shillaker2020faasm,mvondo2021ofc, romero2021faast}. 

To accommodate fluctuating input rates, recent work has explored elastic scaling in serverless stream processing by tuning each operator's degree of parallelism (DoP), i.e., the number of concurrently executing function instances per operator~\cite{li2022faasflow,wen24stepconf,zhang24jolteon}. In stateful serverless environments with distributed scheduling, however, determining the required DoP is only part of the scaling problem. Once the DoP of each operator is chosen, the system must also determine how much workload each worker can sustain and how the resulting instances should be placed across workers. Existing approaches typically model worker capacity using the number of available function-instance slots~\cite{li2022faasflow,nestorov2024dexter}, or leave placement to the underlying platform~\cite{bhasi21kraken,wen24stepconf,zhang24jolteon}. Such approaches account for the capacity consumed by function execution, but overlook the capacity consumed by processing chained requests. Under distributed scheduling, each worker runs a local scheduler alongside its function instances. When an instance completes, the scheduler determines the destination of each chained request and dispatches it accordingly. Requests whose successors are placed on the same worker incur little dispatch overhead, whereas requests sent to remote workers require additional processing such as serialization and network transmission~\cite{yu25pheromone,qi24spright}. We refer to the worker capacity consumed by this scheduling and dispatch activity as the \emph{chained-request cost}. Because this cost is not captured by DoP or slot-based capacity models, a worker may appear to have available execution capacity even when it is already saturated by the combined execution and chained-request workload.

Another question is how many workers to use, which itself involves a trade-off: too few workers cannot sustain the input rate, while spreading the workload across too many forces more chained requests across worker boundaries and inflates end-to-end latency.
Placing operator instances into workers so as to minimize the number of workers, while keeping heavily communicating operators together and respecting each worker's capacity, generalizes the classic operator-placement problem, which is known to be NP-hard~\cite{benoit2013scheduling,adhikari2019schedulingsurvey}. To keep the worker count low, some methods bin-pack instances onto workers according to each worker's available execution slots~\cite{gunasekaran20fifer,li2022faasflow}. Because slot-based methods do not account for the chained-request cost a new placement induces, they pack more workload onto a worker than it can sustain and consequently provision too few workers. Zhang24~\cite{zhang2024efficient} spreads operator instances over more workers than actually required, reducing the workload on each to leave capacity for chained-request cost. However, it controls the spread with user-specified weights and does not quantify the cost of local versus remote chained requests. What is needed is an efficient heuristic method that packs operators onto the fewest workers able to sustain the input rate. Its capacity model must consider execution and chained-request cost together, and must evaluate the chained-request cost under the new placement as operators are assigned.

When scaling the number of workers, operator function instances are migrated across workers, and their co-located state must be redistributed accordingly. Otherwise, state access will frequently cross worker boundaries, leading to high latency~\cite{huang26hydo}. However, existing state-migration schemes are designed for incremental rebalancing rather than for a scaling step. They revisit one operator at a time and relocate part of its state in response to the currently observed imbalance~\cite{wen24lola, Nardelli24offloading}, so a cluster-wide change in placement is reached only after many such rounds. Elastic scaling poses a different requirement. Once a new worker count and operator placement have been selected, the system should transition directly to that placement, relocating the affected operator instances and their state in a coordinated migration step. This allows execution to resume with state co-located at the new placement after a single scaling transition, rather than gradually converging through repeated rebalancing rounds.

To address these challenges, we propose CLASP, a chained-request-aware elastic scaling and operator placement for stream processing applications in stateful serverless environments. CLASP proactively scales the number of workers to accommodate fluctuating input rates while minimizing the number of workers used. To this end, it continuously monitors worker metrics and uses online regression over historical observations to estimate each worker's capacity, accounting for both execution workload and chained-request cost. When scaling is triggered, it packs the projected workloads onto workers, accounting for both execution and chained-request costs. Once the scaling decision is made, it dispatches function instances to their assigned workers and seamlessly migrates the operator state and in-flight requests alongside them, minimizing processing interruption. Implemented on Faasm~\cite{shillaker2020faasm}, our method achieves performance gains over the FaaSFlow~\cite{li2022faasflow} and Zhang24~\cite{zhang2024efficient}, including a throughput improvement of up to 3.3~$\times$ and an end-to-end latency reduction of up to 76\%.

\section{Background and Motivation}
\label{sec:motivation}

\begin{figure*}[htbp]
\centering
\includegraphics[width=0.95\textwidth]{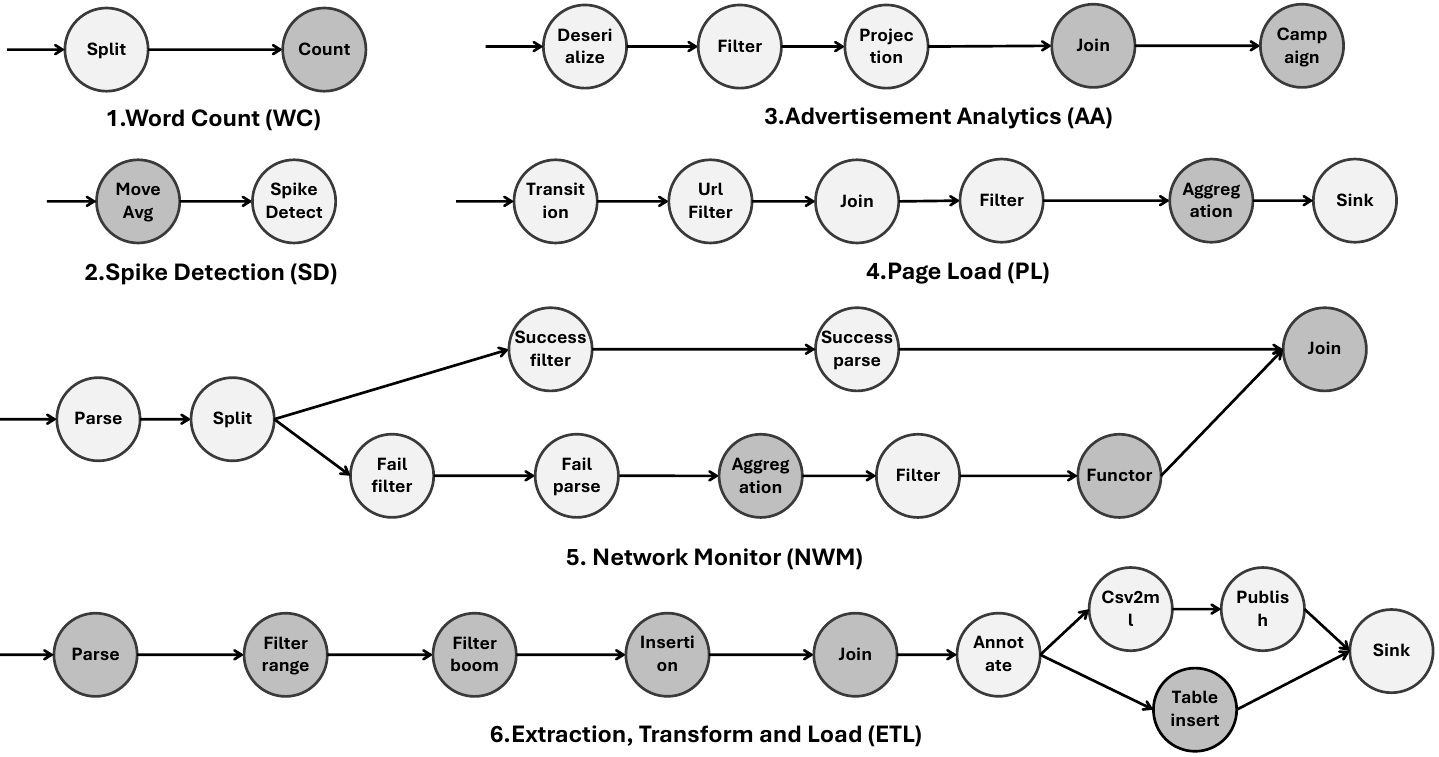}
\caption{Benchmark Application Topologies. Stateless operators are shown in light grey and stateful operators in dark grey.}
\Description{Six directed acyclic graphs, one per benchmark application, drawn as chains of labelled circles connected by arrows. Node shading encodes operator type: light grey for stateless, dark grey for stateful. (1) Word Count: Split feeds Count. (2) Spike Detection: Moving Average feeds Spike Detect. (3) Advertisement Analytics: Deserialize, Filter, and Projection in sequence, then Join, then Campaign. (4) Page Load: Transition, URL Filter, Join, and Filter in sequence, then Aggregation, then Sink. (5) Network Monitor: Parse feeds Split, which forks into two branches. The upper branch runs Success Filter then Success Parse; the lower branch runs Fail Filter, Fail Parse, Aggregation, Filter, and Functor, with several edges. Both branches converge on a single Join. (6) Extraction, Transform and Load: Parse, Filter Range, Filter Bloom, Insertion, and Join in sequence, followed by Annotate, which forks into a Csv2ml then Publish path and a Table Insert path; both converge on Sink.}
\label{fig:3.0.application_topology}
\end{figure*}

In this section, we first present a background on the execution of stream processing applications in serverless environments. We then show that cross-worker chained requests consume non-negligible worker capacity, causing conventional slot-based capacity estimates to overestimate the workload a worker can sustain. Next, we demonstrate that provisioning more workers than necessary increases end-to-end latency, motivating the use of the fewest workers capable of sustaining the input rate. Finally, we show why state must migrate together with operator placement to preserve local state access.

\subsection{Stream Processing in Serverless and Benchmarks}

When running stream processing applications in a serverless environment, each operator is implemented as a function containing the operator's logic. At runtime, the function is instantiated as one or more function instances, which can execute concurrently across workers; the number of such instances is referred to as the operator's parallelism. On each worker, these instances run in a pool of executors, where each executor is a slot that executes one function instance at a time. We refer to the total number of slots a worker provides as its maximum executor slots, which bounds how many instances it can run concurrently. Note that this bound characterizes only function-execution concurrency. As we show in Section~(\ref{sec:bg_distributed_cost}), a worker also spends capacity on scheduling and dispatching chained requests, so its executor-slot count does not by itself determine the load it can sustain. 

Each operator processes incoming tuples. In a serverless setting, a tuple arrives as a request that invokes the operator's function and carries the tuple as its payload. The data flow between operators is realized through chained requests: upon processing a request, a function instance generates one or more chained requests, each invoking the function of a downstream operator. Under distributed scheduling, the local scheduler determines each chained request's destination worker and dispatches it there, consuming different amounts of the worker's capacity depending on whether that destination is local or remote.

Operators in stream processing applications can be broadly classified into two categories: stateless and stateful. A stateless operator processes each incoming request according to its predefined logic without retaining any intermediate information. In contrast, a stateful operator maintains intermediate results (i.e., state) that are acted upon and updated when a request is processed~\cite{carbone17stateflink}. For example, in the count operator of a word count application, the state records how many times each word has been seen so far. In distributed-scheduling stateful serverless environments, each operator is typically bound to a set of workers before any request is routed to it. A local scheduler can therefore dispatch each chained request directly to its destination according to this binding. The state a request acts upon likewise resides on the worker that executes it, so state access stays local~\cite{li2022faasflow, jin2023ditto, zhang2024efficient, wang2025gamethoryscheduling, ghorbian2024schedulesurvey}.

To investigate the performance of stream processing applications in serverless environments, we run experiments on the Faasm~\cite{shillaker2020faasm} stateful serverless platform with a simple bin-packing scheduler based solely on execution workload, and with all state accesses kept local. We evaluate six applications spanning different DAG sizes (2 to 10 operators), topologies (linear and branching), and fractions of executions that access state. Their topologies are shown in Figure~\ref{fig:3.0.application_topology}:

\noindent\textbf{Word Count (WC)} counts the occurrences of each unique word from text sources (2 operators)~\cite{bordin2020dspbench}.

\noindent\textbf{Spike Detection (SD)} maintains a moving window of temperature readings for each sensor ID and uses them to detect whether a newly received event is a spike (2 operators)~\cite{bordin2020dspbench}.

\noindent\textbf{Advertisement Analytics (AA)} maps advertisements to their corresponding campaigns and counts the number of events per campaign within a window (5 operators)~\cite{chintapalli2016yahoobenchmarking}.

\noindent\textbf{Page Load (PL)} computes the average page load time per page category by analyzing a stream of log files (6 operators)~\cite{xu2016stela}.

\noindent\textbf{Network Monitor (NWM)} monitors log files and looks for successful logins that were preceded by multiple failed attempts from the same host (10 operators)~\cite{gedik2014elastic}.

\noindent\textbf{Extraction, Transform and Load (ETL)} processes data streams through filtering, interpolation, and metadata annotation, and then persists the results (10 operators)~\cite{shukla2017riotbench}.

\begin{figure*}[ht]
  \centering
  \begin{minipage}[b]{0.3\textwidth}
    \centering
    \includegraphics[width=\textwidth]{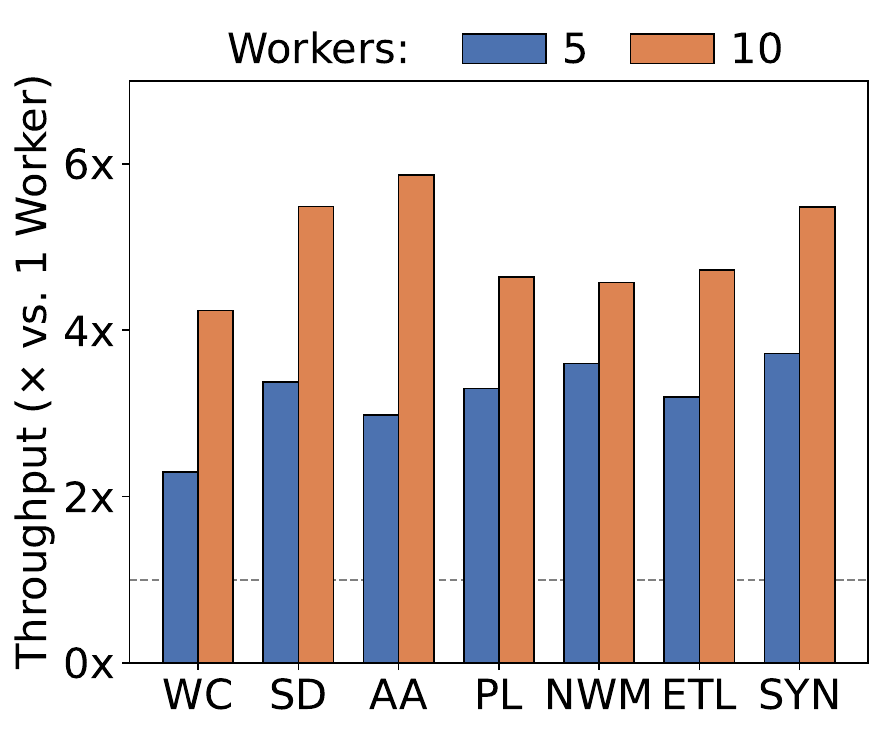}
    \captionof{figure}{Comparison of throughput across cluster sizes (relative to 1 worker).}
    \Description{A grouped bar chart. The horizontal axis lists seven applications: WC, SD, AA, PL, NWM, ETL, and SYN. The vertical axis shows throughput relative to a single worker, from 0x to 6x. Each application has two bars, one for a 5-worker cluster and one for a 10-worker cluster. Every application falls well short of ideal linear scaling: the 5-worker bars sit between roughly 2.5x and 4x rather than 5x, and the 10-worker bars between roughly 4x and 5.5x rather than 10x. The synthetic application SYN, which is evenly balanced by construction, reaches about 3.7x at 5 workers and 5.5x at 10 workers.}
    \label{fig:2.1.bg_throughput}
  \end{minipage}
  \hfill
  \begin{minipage}[b]{0.3\textwidth}
    \centering
    \includegraphics[width=\textwidth]{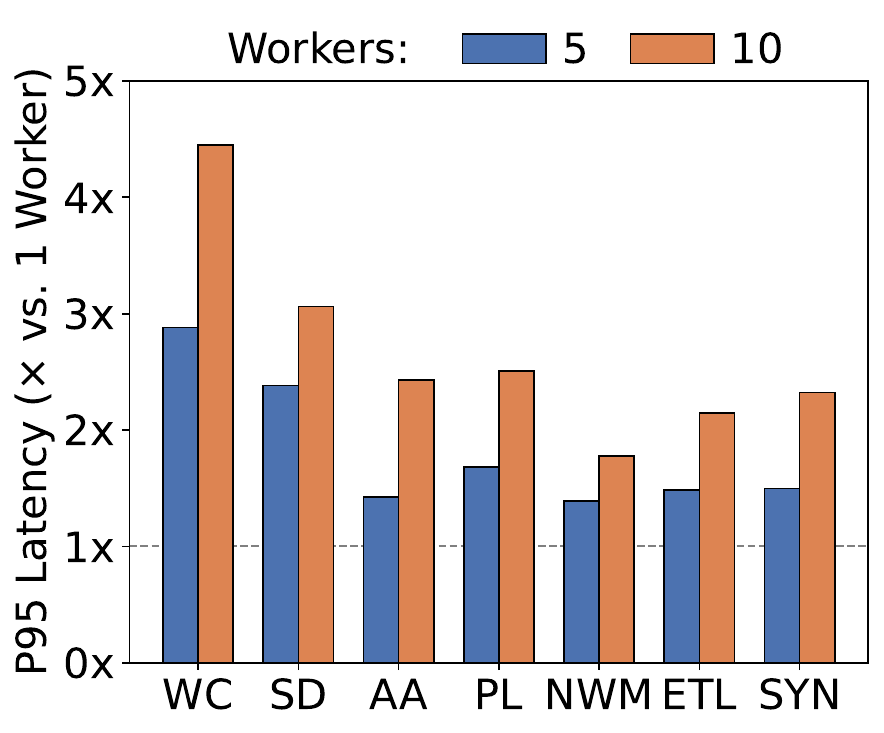}
    \captionof{figure}{Comparison of p95 E2E latency across cluster sizes (relative to 1 worker).}
    \Description{A grouped bar chart with the same seven applications on the horizontal axis as the previous figure: WC, SD, AA, PL, NWM, ETL, and SYN. The vertical axis shows p95 end-to-end latency relative to a single worker, from 0x to over 5x. Each application has two bars, for 5 workers and for 10 workers. In every application the 10-worker bar is taller than the 5-worker bar, so latency rises with cluster size at a fixed input rate. The effect is largest for WC, where latency grows from roughly 2.9x at 5 workers to over 4.4x at 10 workers, and smallest for NWM and ETL, where both bars sit near 2x.}
    \label{fig:2.2.bg_app_latency}
  \end{minipage}
  \hfill
  \begin{minipage}[b]{0.3\textwidth}
    \centering
    \includegraphics[width=\textwidth]{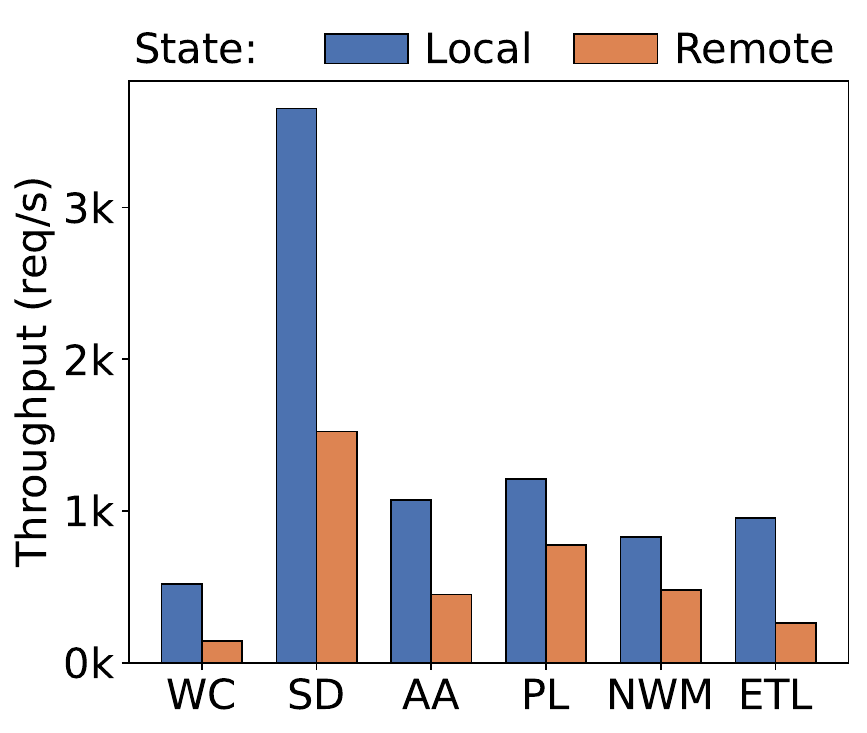}
    \captionof{figure}{Comparison of throughput under different state access circumstances.}
    \Description{A grouped bar chart comparing local and remote state access on a 10-worker cluster. The horizontal axis lists six applications: WC, SD, AA, PL, NWM, and ETL. SYN is omitted because it accesses no state. The vertical axis shows absolute throughput in requests per second, from 0 to about 3000. Each application has a local-access bar and a remote-access bar; the local bar is taller in every case. The gap is widest for SD, whose local bar approaches 3000 while its remote bar sits near 1300, and for WC, where local access yields roughly 3.5 times the throughput of remote access. The gap is narrowest for PL, at roughly 1.5 times, matching its lower rate of state access.}
    \label{fig:2.3.bg_state_latency}
  \end{minipage}
\end{figure*}

\subsection{Chained-Request Cost}
\label{sec:bg_distributed_cost}

Existing approaches first determine each operator's degree of parallelism and then assign the resulting instances to workers based on available slots, overlooking the scheduling and communication overhead that chained requests incur across workers~\cite{li2022faasflow,jin2023ditto,nestorov2024dexter}. To expose this effect, we run the benchmarks in Figure~\ref{fig:3.0.application_topology}, together with a synthetic application SYN, across different numbers of workers (1, 5 and 10) under an unthrottled input stream, and compare the resulting throughput. SYN is a linear pipeline with ten operators, in which each operator implements identical no-op logic and forwards its requests to the next operator in the chain. Due to SYN's uniform structure and identical per-operator logic, it is split evenly across the 1, 5, or 10 workers, giving each worker a similar execution workload and communication cost.

SYN is designed to eliminate the effect of workload imbalance across workers. As shown in Figure~\ref{fig:2.1.bg_throughput}, SYN scales markedly sub-linearly: maximum throughput reaches only about 3.7$\times$ that of a single worker at 5 workers, and 5.5$\times$ at 10 workers. It falls well short of the 5$\times$ and 10$\times$ expected under ideal linear scaling. The other applications exhibit a similar sub-linear trend, though the exact degree varies with each application's pattern. This is because adding workers increases the fraction of chained requests that cross worker boundaries, and remote requests consume more of each worker's capacity than local ones. Therefore, worker capacity cannot be inferred from executor slots or execution workload alone; it must also account for the cost of dispatching chained requests under the current placement.

\subsection{Over-Provisioning Increases Latency}

Under a light input rate, using more workers than required not only wastes resources but also increases end-to-end latency. This is because spreading the application's chained requests across more workers forces more of them to be routed between workers, and such inter-worker communication is more costly and time-consuming than local communication~\cite{carver2020wukong,li2026iroute}. To demonstrate this, we adopt the same benchmarks and experimental setup as in Section~(\ref{sec:bg_distributed_cost}). For each application, the input rate is fixed at the maximum throughput a single worker can sustain, and we measure the resulting end-to-end latency.

As shown in Figure~\ref{fig:2.2.bg_app_latency}, the p95 end-to-end latency is consistently lower on the smaller cluster across all applications, as fewer requests cross worker boundaries. 
For SYN, the p95 end-to-end latency at 10 and 5 workers is 2.3$\times$ and 1.5$\times$ that of a single worker, respectively. The effect is most pronounced in WC and SD, which have only two operators each: with little execution time in the pipeline, the added latency of crossing a worker boundary accounts for a larger share of the end-to-end latency.
This experiment thus highlights the benefit of using the fewest workers that can still sustain the input rate.

\subsection{State Access Overhead}

When handling a request for a stateful operator in a serverless setting, the executor must retrieve the associated state from storage and, if necessary, apply updates to it. Consequently, both the end-to-end latency of executing a request and the resulting maximum throughput are highly sensitive to the state access time.

To expose this effect, we run two baselines with the same bin-packing scheduling strategy but different state-access schemes. In the local-access baseline, an executor retrieves state from the state server on its own worker; in the remote-access baseline, it always retrieves state from the state server on the planner. With the same experimental setup as in Section~(\ref{sec:bg_distributed_cost}), we run the benchmarks in Figure~\ref{fig:3.0.application_topology} on 10 workers under an unthrottled input stream and compare the resulting throughput. SYN is not included here because it does not access any state.

The throughput gap stems from a difference in per-request execution latency. We define the execution latency of an operator as the time from when its executor starts until it completes. In our experimental setup, our measurements show that a stateless operator and a stateful operator with local state access both have an execution latency of around 400 $\mu$s. However, this latency increases to about 1400 $\mu$s for a stateful operator when it accesses the state from a remote server. As shown in Figure~\ref{fig:2.3.bg_state_latency}, the local-access baseline consistently achieves higher throughput than the remote-access baseline, and the gap widens for applications with more frequent state access. Local access yields about a 3.5$\times$ throughput improvement for WC, where 90\% of function executions require state access, but only about 1.5$\times$ for PL, where the figure is 20\%. This result highlights the importance of migrating state along with the new placement, ensuring that stateful operators retain local state access after replacement.

\section{Related Work}

\begin{table*}[ht]
\centering
\renewcommand{\arraystretch}{1.2}
\caption{A comparison with related work.}
\label{tab:related_comparison}
\begin{tabular}{llccccccc}
\hline
Representative Work & Scaling & Concurrent Exec. & OP & ECM & RM & WM & SL & SM \\
\hline
LatEst~\cite{Sfakianakis22latest}         & Vertical   &           &           &           & \ding{51} &           &           &           \\
Ditto~\cite{jin2023ditto}                 & Horizontal &           & \ding{51} & \ding{51} &           &           & \ding{51} &           \\
Dexter~\cite{nestorov2024dexter}          & Horizontal &           &           &           & \ding{51} &           &           &           \\
Jolteon~\cite{zhang24jolteon}             & Horizontal &           &           & \ding{51} &           &           &           &           \\
Kraken~\cite{bhasi21kraken}               & Horizontal & \ding{51} &           &           &           & \ding{51} &           &           \\
StepConf~\cite{wen24stepconf}             & Horizontal & \ding{51} &           &           & \ding{51} &           &           &           \\
Fifer~\cite{gunasekaran20fifer}           & Horizontal & \ding{51} & \ding{51} &           &           & \ding{51} &           &           \\
FaaSFlow~\cite{li2022faasflow}            & Horizontal & \ding{51} & \ding{51} &           &           & \ding{51} & \ding{51} &           \\
Zhang24~\cite{zhang2024efficient}         & Horizontal & \ding{51} & \ding{51} & \ding{51} & \ding{51} &           & \ding{51} &           \\
\hline
\textbf{CLASP (Ours)}                     & Horizontal & \ding{51} & \ding{51} & \ding{51} & \ding{51} & \ding{51} & \ding{51} & \ding{51} \\
\hline
\end{tabular}
\\[1ex]
{\footnotesize \textbf{Concurrent Exec.}: all operators run concurrently vs.\ stage-by-stage;\quad
\textbf{OP}: Operator-to-Worker Placement;\quad
\textbf{ECM}: Execution and Chained-Request Cost Model;\quad
\textbf{RM}: Runtime Cost Estimation from Observations;\quad
\textbf{WM}: Worker Minimization;\quad
\textbf{SL}: State-Locality Preserving (i.e., execute each request on the worker that holds the state it acts upon);\quad
\textbf{SM}: State Migration}
\end{table*}

\paragraph{Vertical Resource Provisioning}
Some research focuses on vertically provisioning resources to a single function instance~\cite{enes2020rtsforBD, Sfakianakis22latest}. LatEst~\cite{Sfakianakis22latest} uses cgroups to control a serverless instance's CPU allocation, employing a feedback controller to estimate the resources needed to meet a target tail latency as the incoming load changes. Such vertical tuning is orthogonal to our setting: with each worker's capacity fixed, sustaining a higher input rate requires adjusting how many workers are used and where operators sit on them, which is what CLASP addresses.

\paragraph{Parallelism Tuning for Batch Processing}
Research has also explored the optimal DoP for operators in batch processing applications, aiming to reduce job completion time or the number of workers used~\cite{yue24demeter,jin2023ditto,nestorov2024dexter,zhang24jolteon}. Ditto~\cite{jin2023ditto} targets batch analytics jobs and jointly optimizes the degree of parallelism and function placement using an execution time model that accounts for both computation and data-transfer time, thereby reducing job completion time and cost. Dexter~\cite{nestorov2024dexter} dynamically adjusts the number of executors at per-operator granularity using a combination of historical prediction and hill climbing search to balance performance and cost. Jolteon~\cite{zhang24jolteon} exploits the convexity of the problem to solve it with gradient descent, finding the resource configuration that minimizes cost or latency under a user-specified bound. However, batch processing executes operators stage by stage, completing the entire batch at one operator before moving to the next, whereas stream processing runs all operators concurrently on continuously arriving data. Because of this, batch processing can keep each operator's chained requests on the worker that produced it, launching the next operator there once the current stage finishes so that the chained requests are executed locally. In stream processing, downstream operators are already running elsewhere when a chained request is produced, so the request must be routed to them right away rather than held for a later stage.

\paragraph{Configuration Tuning with Platform-Managed Placement}
Some existing works jointly optimize the parallelism and resource configuration of each operator, while leaving instance placement to the underlying serverless platform~\cite{bhasi21kraken,wen24stepconf}. StepConf~\cite{wen24stepconf} dynamically configures each function instance at runtime, jointly optimizing memory size and function parallelism to minimize cost under an end-to-end SLO. Because placement is delegated to the platform, these approaches cannot evaluate how a candidate placement changes local and remote chained-request costs, nor can they explicitly minimize the number of workers used.

\paragraph{Configuration Tuning with Self-Managed Placement}
Other research first determines the degree of parallelism of each operator, then co-locates heavily-communicating operators into the same worker according to the available worker resource to reduce latency~\cite{gunasekaran20fifer,li2022faasflow,zhang2024efficient}. FaaSFlow~\cite{li2022faasflow} profiles each operator's instance number from auto-scaling whenever the workflow suffers performance degradation or QoS violation, and adopts a critical-path-based greedy grouping strategy that collocates heavily-communicating operators onto fewest workers with sufficient capacity. However, it measures a worker's capacity only by the number of executors on it, overlooking the chained-request cost and thus causing resource imbalance across workers. Zhang24~\cite{zhang2024efficient} predicts each operator's parallelism using a non-linear performance model that accounts for scheduling overhead and resource contention, and then greedily assigns the resulting instances to workers by scoring both data affinity and load balancing. The data-affinity term reduces remote chained-request transfer, while the load-balancing term spreads executor instances across workers to exploit the available resources of the cluster. However, it does not quantify the cost of local versus remote chained requests, and instead relies on a user-specified weighting between the two scoring terms. CLASP differs from these approaches by estimating execution and chained-request costs directly from runtime observations, and by accounting for both costs as assigning operators into workers, with the aim of minimizing the number of workers used while sustaining the input rate.

\paragraph{State Migration in Stateful Serverless}
Existing state-migration schemes treat relocation as incremental rebalancing rather than as part of a scaling step: they periodically re-examine the current load distribution and move part of a single operator's state in response to the imbalance they observe~\cite{wen24lola,Nardelli24offloading}. LoLa~\cite{wen24lola} migrates operator state whenever workload imbalance across workers exceeds a threshold; each trigger relocates part of a single operator's state, so a cluster-wide realignment is reached only after many rounds. CLASP instead derives a single migration plan from the difference between the old and new placements, moving every operator's complete state together with its instances in one round.

\section{CLASP Design}

\begin{figure}[ht]
\centering
\includegraphics[width=0.65\textwidth]{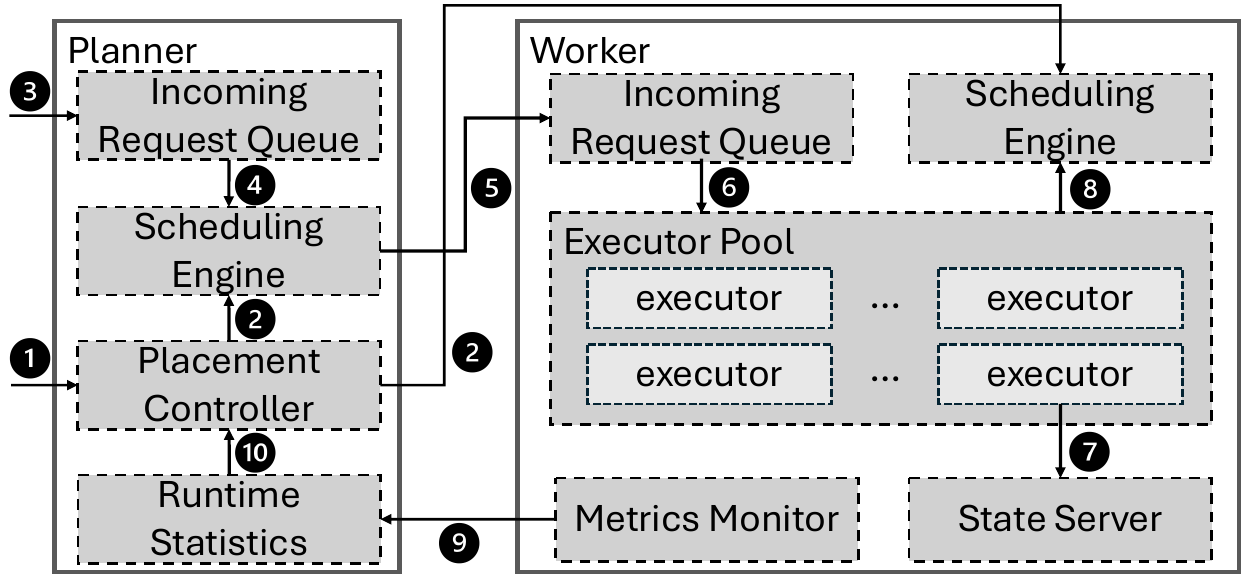}
\caption{Architecture of the designed method.}
\label{fig:3.architecture}
\Description{A block diagram with two boxes side by side: a planner on the left and a worker on the right, representing one of several identical workers. The planner box contains four stacked components, from top to bottom: an incoming request queue, a scheduling engine, a placement controller, and a runtime statistics module. The worker box contains an incoming request queue and a scheduling engine along the top, an executor pool in the middle holding four executor slots, and a metrics monitor and a state server along the bottom. Ten numbered arrows trace the flow: the user submits the application to the placement controller (1), which distributes the computed placement to the scheduling engines on both the planner and the worker (2); input requests enter the planner's queue (3), are fetched by its scheduling engine (4), and are dispatched to the worker's incoming queue (5); the executor pool fetches and executes them (6), reading state from the local state server when the operator is stateful (7); on completion, executors hand the resulting chained requests to the worker's scheduling engine (8); the metrics monitor periodically reports to the planner's runtime statistics module (9); and the placement controller uses those metrics to rescale and update the placement (10).}
\end{figure}

CLASP operates as a continuous monitor–predict–place–migrate cycle. It first collects runtime metrics and estimates the effective capacity of each worker. When the observed input rate changes sufficiently, it projects the application load onto this capacity, computes a new operator placement, derives the parallelism each operator requires, and migrates operator state together with queued requests to the selected workers.

Figure~\ref{fig:3.architecture} illustrates the architecture that realizes this cycle. The system consists of a single planner and multiple workers, each running independently on its own host. We assume a homogeneous cluster, in which every worker provides the same capacity. The planner runs a placement controller that receives the application from the user~\filledcircled{1}, computes an initial placement of operators onto workers, and distributes it to the scheduling engines on both the planner and the workers~\filledcircled{2}. After initialization, input requests enter the planner's incoming request queue~\filledcircled{3}; the scheduling engine fetches them~\filledcircled{4}, selects the destination worker for each request according to the operator placement, and dispatches it to that worker's incoming request queue~\filledcircled{5}. On each worker, the executor pool fetches the requests from its incoming queue and executes them~\filledcircled{6}. During the execution of a stateful operator, the executor reads the required state from the local state server~\filledcircled{7}. Once execution completes, the executor forwards the resulting chained requests to the local scheduling engine~\filledcircled{8}, which dispatches them to the appropriate workers according to the operator placement. A metrics monitor on each worker records runtime metrics and periodically reports them to the runtime statistics module in the planner~\filledcircled{9}. Based on these metrics, the scheduling engine scales the number of workers and adjusts the operator-to-worker placement~\filledcircled{10}.

\subsection{Metrics Monitor}

To support scaling decisions, the system collects runtime metrics from both the workers and the planner. Each worker monitors its incoming request queue length, the number and latency of executed requests, the chained-request rate between operators, and the rate of chained requests it generates to local and remote workers. The planner then aggregates these per-worker statistics and additionally tracks the length of its own incoming request queue, as well as the system-wide input rate and throughput. All monitored metrics are summarized in Table~\ref{tab:metrics}.

\begin{table}[ht]
\centering
\caption{Runtime metrics monitored. Rates are measured per second.}
\label{tab:metrics}
\begin{tabular}{cp{7cm}}
\hline
\textbf{Symbol} & \textbf{Description} \\
\hline
\multicolumn{2}{l}{\textit{Worker} $w$} \\ 
$Q_w$          & Size of worker $w$'s incoming request queue \\
$R^l_w$        & Rate of local chained requests generated on worker $w$  \\
$R^r_w$        & Rate of remote chained requests generated on worker $w$  \\
$\delta_w$     & Number of distinct remote workers reached by worker $w$'s chained requests \\
$N_{w,o}$      & Rate of processed requests of operator $o$ on worker $w$\\
$\bar{\tau}_{w,o}$ & Average request execution time of operator $o$ on worker $w$ ($\mu$s)\\
$E^{o\to o_s}_{w}$ & Rate of chained requests generated on worker $w$ from operator $o$ to its successor $o_s$ \\
\hline
\multicolumn{2}{l}{\textit{Planner} $p$} \\
$Q_p$          & Size of the planner's incoming request queue \\
$\lambda_p$    & System input rate \\
$\Lambda$    & System throughput \\
\hline
\end{tabular}
\end{table}

\subsection{Worker Capacity Model}
\label{sec:worker_capacity_model}

To model a worker's load, we express its total load as the sum of execution and chained-request overheads, where the latter is further broken down into local chained-request rate, remote chained-request rate, and the number of destination workers:

\begin{equation}
    W_w = \sum_{o} N_{w,o} \cdot \bar{\tau}_{w,o} + \alpha \cdot R^l_w + \beta \cdot R^r_w + \gamma \cdot \delta_w
\label{eqt:3.1.workercapacity}
\end{equation}

where $\sum_{o} N_{w,o} \cdot \bar{\tau}_{w,o}$ captures the total execution cost across all operators in $w$, and $\alpha$, $\beta$, $\gamma$ are coefficients weighting the overhead of local chained-request dispatching, remote chained-request dispatching, and the number of distinct remote destination workers, respectively. Since these coefficients may vary across deployments, we estimate them online from runtime observations, as described in the following section.

\subsubsection{Online Coefficient Estimation}

The scaling policy requires estimates of the maximum worker capacity $W$ and the overhead coefficients $\alpha,\beta,\gamma$ of Equation~(\ref{eqt:3.1.workercapacity}). None can be measured directly, so we estimate them online from \emph{saturated} workers---those whose incoming queue $Q_w$ grows persistently (e.g., over three consecutive seconds) or whose queuing delay exceeds a threshold. When a worker saturates, its arrival rate exceeds the rate at which it can process requests; the load it actually sustains, $W_w$, then equals its maximum capacity, i.e.\ $W_w\!=\!W$. Since the cluster is homogeneous, $W,\alpha,\beta,\gamma$ are shared across workers, so saturated samples from all workers are pooled into a single online regression, raising the effective sampling rate and averaging out per-worker noise.

\paragraph{Estimating the coefficients}
Writing the aggregate execution cost as $E_w=\sum_{o} N_{w,o}\,\bar{\tau}_{w,o}$, a saturated worker satisfies $E_w = W-\alpha R^l_w-\beta R^r_w-\gamma\delta_w$. Differencing two consecutive saturated observations cancels the constant $W$ and leaves one linear constraint on the coefficients,
\begin{equation}
\Delta E^{(k)} = -\,\alpha\,\Delta R^{l,(k)}
                 -\,\beta\,\Delta R^{r,(k)}
                 -\,\gamma\,\Delta \delta^{(k)} + \varepsilon^{(k)},
\label{eqt:diff}
\end{equation}
where $\varepsilon^{(k)}$ absorbs measurement noise and slow drift in $W$, and $\Delta(\cdot)^{(k)}=(\cdot)^{(k)}-(\cdot)^{(k-1)}$. Since $R^l_w$, $R^r_w$, and $\delta_w$ differ by orders of magnitude in their typical values, we normalize each variable by a fixed per-term scaling constant reflecting its magnitude before forming the regression, and rescale the recovered coefficients back to their original units afterward. We then solve for $\boldsymbol{\theta}=[\alpha,\beta,\gamma]^{\top}$ via recursive least squares (RLS) with forgetting factor $\rho\in(0,1)$: each observation updates $\boldsymbol{\theta}$ in $O(1)$ time, and the estimator tracks drift across workload and deployment changes.

\paragraph{Recovering the Maximum Worker Capacity}
Differencing identifies $\alpha,\beta,\gamma$ but eliminates $W$, which is required by the scaling policy. Once $\boldsymbol{\theta}$ is fixed, each saturated observation directly recovers a single-observation estimate $\widehat{W}$ of the maximum worker capacity,
\begin{equation}
\widehat{W}^{(k)} = E_w^{(k)}
  + \alpha\,R^{l,(k)}_w + \beta\,R^{r,(k)}_w + \gamma\,\delta^{(k)}_w.
\label{eqt:worker_W_estimate}
\end{equation}
To suppress per-observation noise, we smooth these estimates with an exponential moving average, yielding the smoothed maximum worker capacity $\bar{W}$:
\begin{equation}
\bar{W}\leftarrow(1-\eta)\,\bar{W}+\eta\,\widehat{W}^{(k)}.
\label{eqt:worker_smooth}
\end{equation}
This smoothed value $\bar{W}$ is then used when deciding the number of workers.

\subsection{Worker Scaling and Operator Scaling}
\label{sec:worker_scaling_policy}

During scaling, the choice of worker count must balance two competing pressures (\S\ref{sec:motivation}): too few workers saturate under load and inflate end-to-end latency, whereas too many incur additional inter-worker communication overhead and waste resources. Using the monitored metrics and the estimated maximum worker capacity $\bar{W}$, the planner continuously checks whether the worker pool can sustain the current input rate, and scales out or in to use the fewest workers that suffice.

Rather than determining each operator's degree of parallelism (DoP) independently, our method assigns operators to workers based on their processing cost and communication pattern, and then derives each operator's parallelism from the resulting placement.

\subsubsection{Load Derivation}
As shown in Table~\ref{tab:metrics}, each worker reports the per-operator processed counts $N_{w,o}$, the average execution times $\bar{\tau}_{w,o}$, and the inter-operator chained-request rates $E^{o\to o_s}_{w}$. In the planner, the placement controller aggregates these across workers into each operator's total processed rate $N_o$, average execution time $\bar{\tau}_o$, and per-edge request rate $E^{o\to o_s}$.

These metrics are observed while the system sustains a \emph{source rate} $\Lambda$ (the measured throughput), whereas placement must be computed for a \emph{target rate} $\lambda_p$ (the current input rate the system must sustain). We therefore project the observed rates from the source rate to the target rate by the scaling factor
\begin{equation}
s = \frac{\lambda_p}{\Lambda},
\end{equation}
giving
\begin{equation}
\tilde{N}_o = s\,N_o, \qquad \tilde{E}^{o\to o_s} = s\,E^{o\to o_s}.
\end{equation}
The average execution time $\bar{\tau}_o$ is taken from the observed metrics and left unscaled. Each operator's projected processing demand is then
\begin{equation}
\tilde{D}_o = \tilde{N}_o\,\bar{\tau}_o,
\end{equation}
which, together with the projected chained-request rates $\tilde{E}^{o\to o_s}$, forms the input to the placement in the next section.

\subsubsection{Reverse-Topology Operator Placement}
\label{sec:reverse_placement}

With the projected load, we place operators onto workers via bin-packing, allowing the number of workers used to emerge from the packing. For an operator, the workload consists of independent requests that can be executed on different workers, so its demand $\tilde{D}_o$ may be split across workers.

We represent a placement as $\Phi=\{\phi_{o,w}\}$. Here, $\phi_{o,w}\ge 0$ is the share of $\tilde{D}_o$ assigned to worker $w$, so that $\sum_w \phi_{o,w}=\tilde{D}_o$. The fraction of operator $o$ placed on worker $w$ is then $\sigma_o(w)=\phi_{o,w}/\tilde{D}_o$. Following the observed routing pattern, a chained-request invocation from $o$ on $w$ to its successor $o_s$ stays local with probability $\sigma_{o_s}(w)$ and goes remote otherwise.

Substituting into the capacity model of Equation~(\ref{eqt:3.1.workercapacity}), the load of worker $w$ under placement $\Phi$ is
\begin{equation}
\begin{split}
L_w(\Phi)={}&\sum_o \phi_{o,w} +\gamma\,\delta_w(\Phi)\\
&+\sum_{(o\to o_s)}\tilde{E}^{o\to o_s}\,\sigma_o(w)\\
&\quad\times\big[\alpha\,\sigma_{o_s}(w)+\beta\big(1-\sigma_{o_s}(w)\big)\big],
\end{split}
\end{equation}
where $\delta_w(\Phi)$ counts the distinct remote workers that $w$'s outgoing chained requests reach.

Guided by this load model, we place operators onto workers subject to the maximum worker capacity, $L_w(\Phi)\le\bar{W}$ for all $w$. The bin-packing co-locates heavily communicating operators to reduce remote chained requests, while using as few workers as possible.

\paragraph{Reverse-topological order}
When placing an operator on a worker, the transmission cost of its chained requests is unknown, since its successor operators may not yet be placed. Visiting operators in \emph{reverse-topological} order removes this dependency: each operator is placed only after all its successors, so this cost is known at the time of placement. We realize this order with a depth-first traversal starting from the sink operators. At each operator, the traversal prioritizes the outgoing edge with the highest chained-request rate, encouraging operators with intensive communication to be placed together. An operator is placed only after all its successors have been placed. If a successor is still unplaced, we skip the operator for now, place the remaining successors first, and return to it afterward.

\begin{figure}[ht]
\centering
\includegraphics[width=0.3\textwidth]{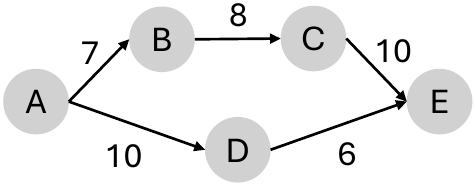}
\caption{DAG of an example application; the number on each edge denotes its chained-request rate.}
\Description{A small directed acyclic graph with five circular nodes labelled A through E. Node A has two outgoing edges: one to B with a chained-request rate of 7, and one to D with a rate of 10. B has a single edge to C with a rate of 8, and C has an edge to E with a rate of 10. D also has an edge to E, with a rate of 6. E is the only sink. The graph is used to illustrate the reverse-topological traversal order, which places E first, then C and B, then D, and finally A.}
\label{fig:3.4.placement_example}
\end{figure}

For the application in Figure~\ref{fig:3.4.placement_example}, the placement starts from $E$, followed by $C$ and $B$. When the traversal reaches $A$, it is postponed because its successor $D$ is not yet placed, and the algorithm proceeds to $D$ instead. Once $D$ is placed, $A$ is finally placed as well.

\paragraph{Per-operator placement}

On our homogeneous cluster, where every worker has the same capacity $\bar{W}$, we place operators using a next-fit bin-packing strategy: an operator fills the current worker, and a new worker is opened only when the current one cannot hold its remaining demand, so the placement uses as few as this heuristic yields.

We place operator $o$ into worker $w$'s remaining capacity $w_{\text{rem}}$. When a worker is first opened, $w_{\text{rem}}=\bar{W}$; each placed operator then consumes part of it. Since $o$'s successors are already placed, we know which workers its chained requests reach, which gives the fan-out cost $\gamma\,g_w$ of placing operator $o$ on $w$. Here, $g_w$ is the number of \emph{new} remote workers that $o$ adds to $w$'s existing connections; remote workers that $w$ already connected to through previously placed operators incur no extra fan-out cost. The incremental load incurred by assigning one unit of operator $o$'s demand to worker $w$ is $m_w=1+\alpha\,\sigma_{o_s}(w)+\beta\big(1-\sigma_{o_s}(w)\big)$. The maximum amount of $o$ that fits on $w$ then follows directly:
\begin{equation}
\phi^{\max}_{o,w}=\frac{w_{rem}-\gamma\,g_w}{m_w}.
\label{eqt:place-maxfit}
\end{equation}
We then update $w$'s remaining capacity accordingly, opening a new worker if $o$'s demand is not fully placed. If the cluster has already reached its worker limit, we stop opening new workers and instead use the placement that sustains the highest throughput on the workers we have. We find this placement with a binary search over the input rate, using the maximum throughput of a single worker multiplied by the number of available workers as the initial upper bound. The search terminates once the interval narrows below a coarse tolerance (e.g., 50 requests/s). This placement needs to be computed only once and is cached for reuse whenever the cluster reaches its worker limit again.

\paragraph{Parallelism}

Once the placement $\Phi$ is fixed, each operator's parallelism $P_o$ (its number of parallel executing function instances) is derived from it. Let $P_{\text{tot}}^{\text{cur}}$ and $P_{\text{tot}}$ denote the current and projected total numbers of instances across all operators. CLASP first scales this total by the scaling factor $s$ used for the load derivation,
\begin{equation}
P_{\text{tot}} = s \cdot P_{\text{tot}}^{\text{cur}}.
\end{equation}
It then allocates the projected instances across operators in proportion to each operator's processing demand,
\begin{equation}
P_o = P_{\text{tot}} \cdot \frac{N_o\,\bar{\tau}_o}{\sum_{o'} N_{o'}\,\bar{\tau}_{o'}}.
\end{equation}
Finally, each operator's $P_o$ instances are distributed across workers according to its allocated shares $\phi_{o,w}$.

In the initial placement, all operators run on a single worker, and all executor slots $P_{\max}$ on the worker are allocated across operators in proportion to each operator's share of the processing demand (observed during the warm-up phase),
\begin{equation}
P_o = P_{\max}\cdot\frac{N_o\,\bar{\tau}_o}{\sum_{o'}N_{o'}\,\bar{\tau}_{o'}}.
\end{equation}

\subsection{State Migration}
\label{sec:state_migration}

\begin{figure}[ht]
\centering
\includegraphics[width=0.65\textwidth]{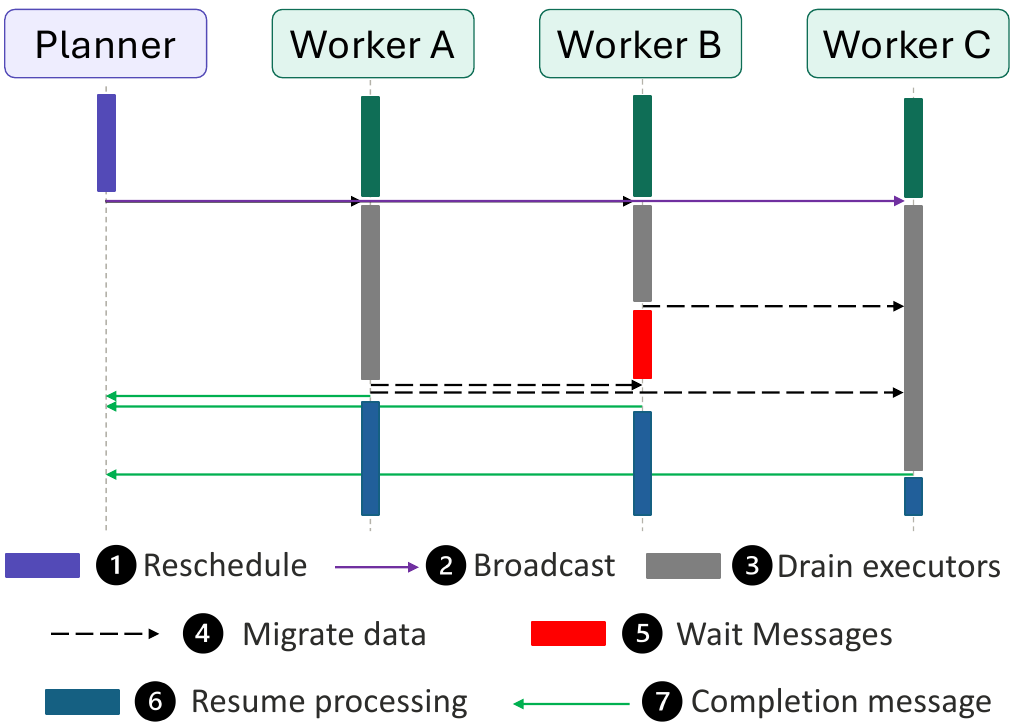}
\caption{State migration sequence with three workers. \textbf{Setup:} the new placement requires $A$ to send data to $B$ and $C$, and $B$ to send data to $C$; $C$ only receives. \textbf{Execution:} $A$ receives from no worker, so it resumes immediately after sending its data; $B$ resumes once $A$'s data arrives; and $C$, by the time it finishes draining, has already received data from both $A$ and $B$, so it too resumes without extra waiting. $A$ and $B$ resume before $C$, without a global synchronization barrier.}
\Description{A sequence diagram with four vertical lifelines labelled Planner, Worker A, Worker B, and Worker C, time flowing downward. Seven event types are distinguished by colour and line style, given in a legend below the diagram: a reschedule step on the planner, a broadcast arrow from the planner to all three workers, a drain- executors phase shown as a grey bar on each worker, dashed migrate- data arrows between workers, a wait-messages period shown as a red bar, a resume-processing marker shown as a blue bar, and completion- message arrows from each worker back to the planner. The planner reschedules, then broadcasts the migration plan to all workers, which each begin draining their executors. Worker A sends data to B and C and, having no incoming transfers, resumes immediately. Worker B sends data to C and shows a short red wait bar before A's data arrives, then resumes. Worker C receives from both A and B and finishes draining after both transfers have landed, so it resumes with no wait bar. The resume markers on A and B occur earlier than the one on C, showing that workers restart independently rather than at a common barrier. Each worker then sends a completion message to the planner.}
\label{fig:3.3.migration}
\end{figure}

After a scaling decision, both the state and the unprocessed requests waiting in the incoming request queue must be migrated according to the new placement, so that each operator continues to execute on the worker that holds the state it acts upon. The migration step is illustrated with an example in Figure~\ref{fig:3.3.migration}. First, the placement controller in the planner compares the new placement with the existing one. From this comparison, the planner derives a migration plan that specifies, for each worker, which peers to send state and requests to and which to receive them from~\filledcircled{1}. It then broadcasts this plan to all workers~\filledcircled{2}. Upon receiving the plan, each worker first waits for its currently running executions to complete~\filledcircled{3}. Then, following the new placement, it computes the destination of each state object in its local state server and each request still waiting in its queue, and dispatches them to the corresponding workers~\filledcircled{4}. After dispatching, the worker waits until it has received all the state and requests destined for it~\filledcircled{5}. This wait is local: the migration plan already specifies, for each worker, which peers will send data to it, so a worker simply waits until it has received data from all of them, independently of any other worker. Once this condition is met, the worker resumes normal request processing~\filledcircled{6} and sends the completion message to the planner~\filledcircled{7}.

This design keeps the planner out of the migration itself: it only computes and broadcasts the plan, while all state and request transfers happen directly between workers. Since each worker's resume condition is local, the migration is fully decentralized, with no global synchronization barrier and no central coordinator that could bottleneck the transfers.

\section{Performance Evaluation}
\label{sec:evaluation}

\noindent\textbf{Implementation.} CLASP is built on top of Faasm~\cite{shillaker2020faasm} and extends it with three key features: (i) an incoming request queue on both the planner and each worker, (ii) a distributed scheduling engine on each worker, and (iii) a metrics monitor on each worker. In our system, all chained requests scheduled to the local worker are stored and transferred through shared memory to reduce communication latency. 

\subsection{Experiment Setup}
\subsubsection{Serverless Baselines and Testbed}
All experiments are conducted on an 11-node Kubernetes cluster. One planner
node runs on an 16-core AMD EPYC 2.30 GHz machine with 64 GB of RAM, while each of the ten worker nodes runs on a 8-core AMD EPYC 2.30 GHz machine with 32 GB of RAM. The maximum number of concurrent executors per worker node is set to 10. We compare our method against two baselines. The first is FaaSFlow~\cite{li2022faasflow}. It profiles each operator's parallelism based on its invocation rate, and then greedily packs heavily-communicating operators onto workers with sufficient slots. The second is Zhang24~\cite{zhang2024efficient}, which predicts each operator's parallelism with a non-linear performance model accounting for scheduling overhead and resource contention, and assigns the resulting instances to workers by scoring both data affinity (favoring placements that keep chained requests local rather than remote) and load balancing. Since all methods rely on runtime statistics to evaluate the resources required by each operator and map them effectively to workers, we first run a 10,000-request warm-up phase. Based on the statistics collected, we then remap the operators and begin recording performance. For our method, $\alpha$, $\beta$ and $\gamma$ are updated online at runtime; we initialize them to $38.5$, $275$ and $51500$, the values that the online estimator converges to across applications and cluster sizes in Section~(\ref{sec:exp_coeff}), so that it starts close to its steady state. For Zhang24, the weights of data affinity and load balancing are each set to 0.5. For all methods, state and request migration follows the same procedure as CLASP in Section~(\ref{sec:state_migration}). Scaling is triggered under two conditions: when the input rate changes consistently for 2 seconds, or when the current throughput fails to match the input rate for 3 seconds.

\subsubsection{Evaluation Metrics and Benchmark Applications}
Our primary goal is to sustain the input rate with the minimum number of workers. We define \emph{throughput} as the number of requests processed end-to-end (i.e., by all operators) per second. We also examine the impact on end-to-end latency, where \emph{end-to-end latency} refers to the time it takes for a request and all of its subsequent chained requests to be processed by the entire pipeline, encompassing both scheduling and execution time. Each experiment was run five times, with each run lasting ten minutes. The results presented in this paper are the average of these five runs. 

\subsection{Migration Cost}

\begin{figure*}[t]
\centering
\includegraphics[width=0.95\textwidth]{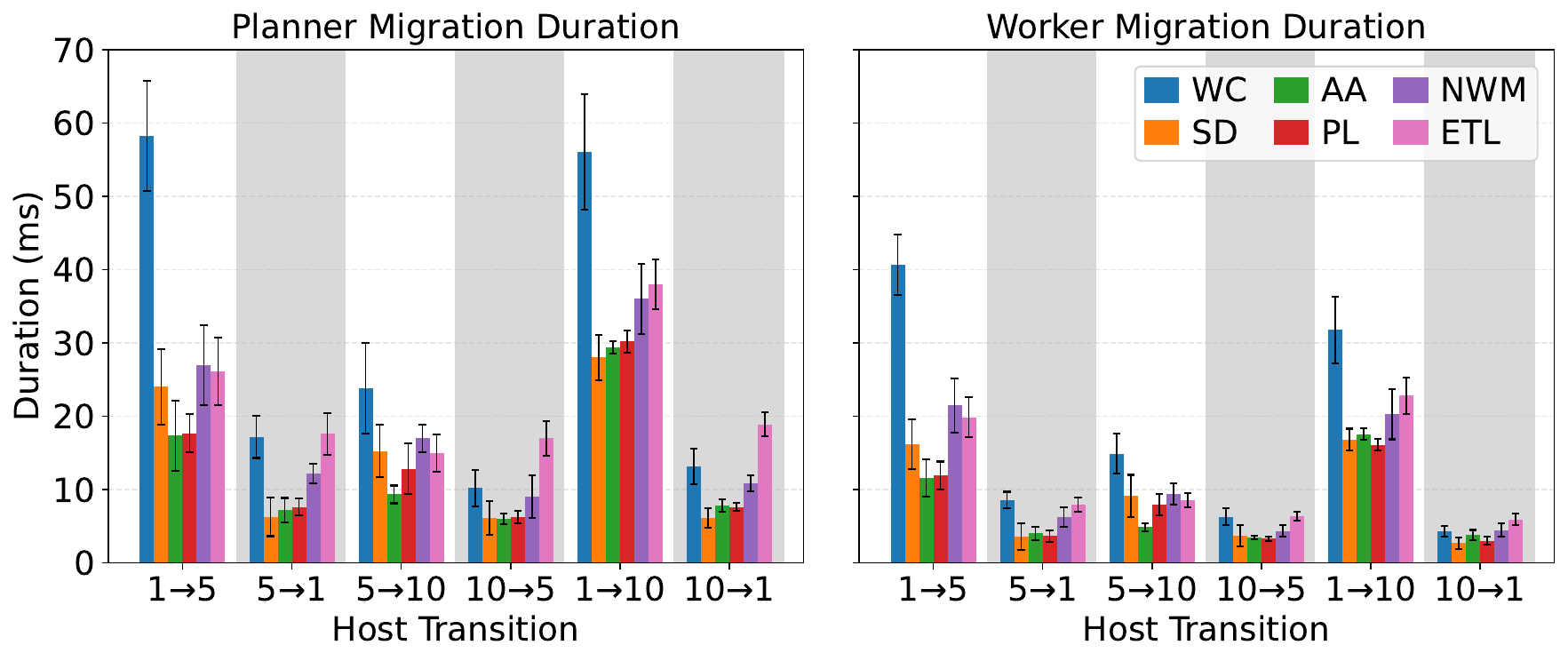}
\caption{Comparison of migration time across different worker transitions and applications.}
\label{fig:4.3.migration}
\Description{Two grouped bar charts side by side, titled Planner Migration Duration and Worker Migration Duration. Both share a horizontal axis of six worker-count transitions: 1 to 5, 5 to 1, 5 to 10, 10 to 5, 1 to 10, and 10 to 1. Alternating grey bands separate the transition pairs. The vertical axis is duration in milliseconds, from 0 to 60. Within each transition, six bars give the six applications WC, SD, AA, PL, NWM, and ETL. In the planner panel, most bars fall between 10 and 20 milliseconds; WC is the tallest in every transition, reaching about 58 milliseconds for 1 to 5 and about 55 for 1 to 10. Scale-up transitions are consistently taller than their scale-down counterparts, and the two transitions that start from a single worker are the tallest of all. The worker panel follows the same pattern at roughly half the magnitude, staying below 10 milliseconds for most applications except when scaling up from a single worker, where WC reaches about 40 milliseconds.}
\end{figure*}

This experiment evaluates the migration cost described in Section~\ref{sec:state_migration}. Before migrating, we run each application for 100 seconds to let the state fill up, and then trigger scaling across six worker transitions: $1\to5$, $5\to10$, $1\to10$, and their reverses. We measure the resulting migration time for each application, distinguishing two durations. The \emph{planner duration} spans from the moment the planner issues the migration plan to the moment it receives completion messages from all workers. The \emph{worker duration} spans from the moment a worker receives the migration plan to the moment it resumes execution.

As shown in Figure~\ref{fig:4.3.migration}, the planner duration is typically between 10 and 20 ms, though three factors make it longer. First, across applications, WC has the longest planner duration. Because it counts word frequencies in articles, WC accumulates a large number of unique words (around 40k at migration time), producing a correspondingly large state and therefore a longer migration. Second, scaling up takes longer than scaling down on average. Although both directions involve the same workers, scaling up spreads data across more destination workers, and each destination must wait until it has received data from all its sources. Third, scaling from a single worker (1$\to$5 and 1$\to$10) takes longer duration than other scaling up transitions. Since all data is stored on one source worker, that worker must dispatch to every destination, and this concentrated transfer volume prolongs the migration.

Across most applications and transitions, workers resume processing within 10 ms, showing that the decentralized migration mechanism introduces only a short execution pause.

\subsection{Online Estimation of the Worker Capacity Model Parameters}
\label{sec:exp_coeff}

\begin{figure*}[t]
\centering
\includegraphics[width=0.9\textwidth]{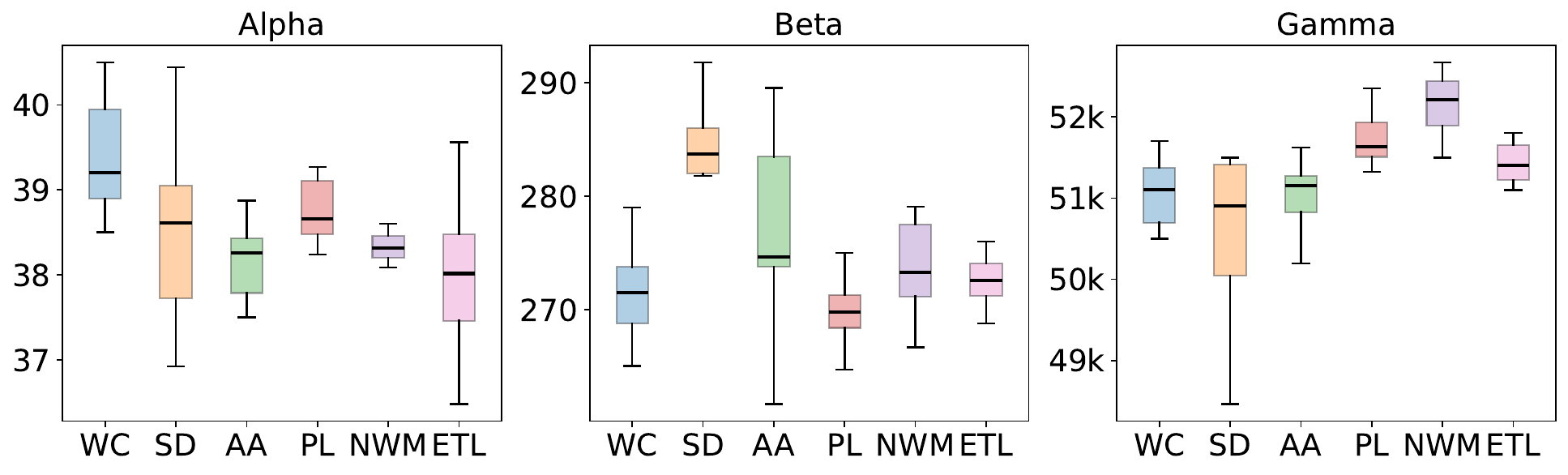}
\caption{Estimated coefficients $\alpha$, $\beta$ and $\gamma$ for each application. Each box aggregates the converged estimates obtained under worker counts from 1 to 10.}
\label{fig:4.3.parameters}
\Description{Three box plots side by side, titled Alpha, Beta, and Gamma, showing the distribution of each estimated coefficient across worker counts from 1 to 10. All three share a horizontal axis listing six applications: WC, SD, AA, PL, NWM, and ETL. The Alpha panel spans roughly 37 to 40 with a median near 38.5; the Beta panel spans roughly 265 to 292 with a median near 275; the Gamma panel spans roughly 49,000 to 52,500 with a median near 51,500. In each panel the boxes for the six applications overlap heavily and the spread within each box is small relative to the coefficient's magnitude, indicating that the online estimator converges to similar values regardless of application and worker count.}
\end{figure*}

In this section, we estimate the coefficients $\alpha$, $\beta$, and $\gamma$ in Equation~(\ref{eqt:3.1.workercapacity}) under different experimental conditions. To do so, we run each application from 1 to 10 workers under an unthrottled input rate and observe the coefficients to which the online estimator converges. Because $R^l_w$, $R^r_w$, and $\delta_w$ differ by orders of magnitude in their typical value, the regression operates on scaled variables; in this scaled space, we initialize $\alpha$, $\beta$, and $\gamma$ to 100, 1000, and 100000, respectively, and update them online from saturated observations via recursive least squares. 

As shown in Figure~\ref{fig:4.3.parameters}, each subplot shows the distribution of the parameter estimates in different applications. The coefficients estimated by the online linear regression average around $\alpha=38.5$, $\beta=275$, and $\gamma=51500$, while the average execution time of an operator in our experiments is around 400$\mu$s.

These values reveal the relative cost of executing a request versus dispatching its chained requests, i.e., scheduling them locally and routing them to their destination workers. Since all terms in Equation~(\ref{eqt:3.1.workercapacity}) contribute to the same worker load, the coefficients are directly comparable: dispatching a remote chained request ($\beta=275$) costs slightly less than executing a request ($\sim$400\,$\mu$s on average). A local chained request ($\alpha=38.5$), by comparison, costs only about one-seventh as much as a remote one, reflecting the overhead of crossing worker boundaries. Opening a chained request to one additional remote worker ($\gamma=51500$) is roughly as expensive as executing 130 requests, making fan-out the dominant cost when an operator's chained requests are scattered across many workers. In our environment, the smoothed maximum worker capacity $\bar{W}$ estimated is around 850000, estimated via Equation~(\ref{eqt:worker_W_estimate})-(\ref{eqt:worker_smooth}), meaning a worker can run around 2000 requests locally per second without any communication cost.

\subsection{Operator Placement}
\label{sec:exp_operator_placement}

\begin{figure}[t]
\centering
\includegraphics[width=0.55\textwidth]{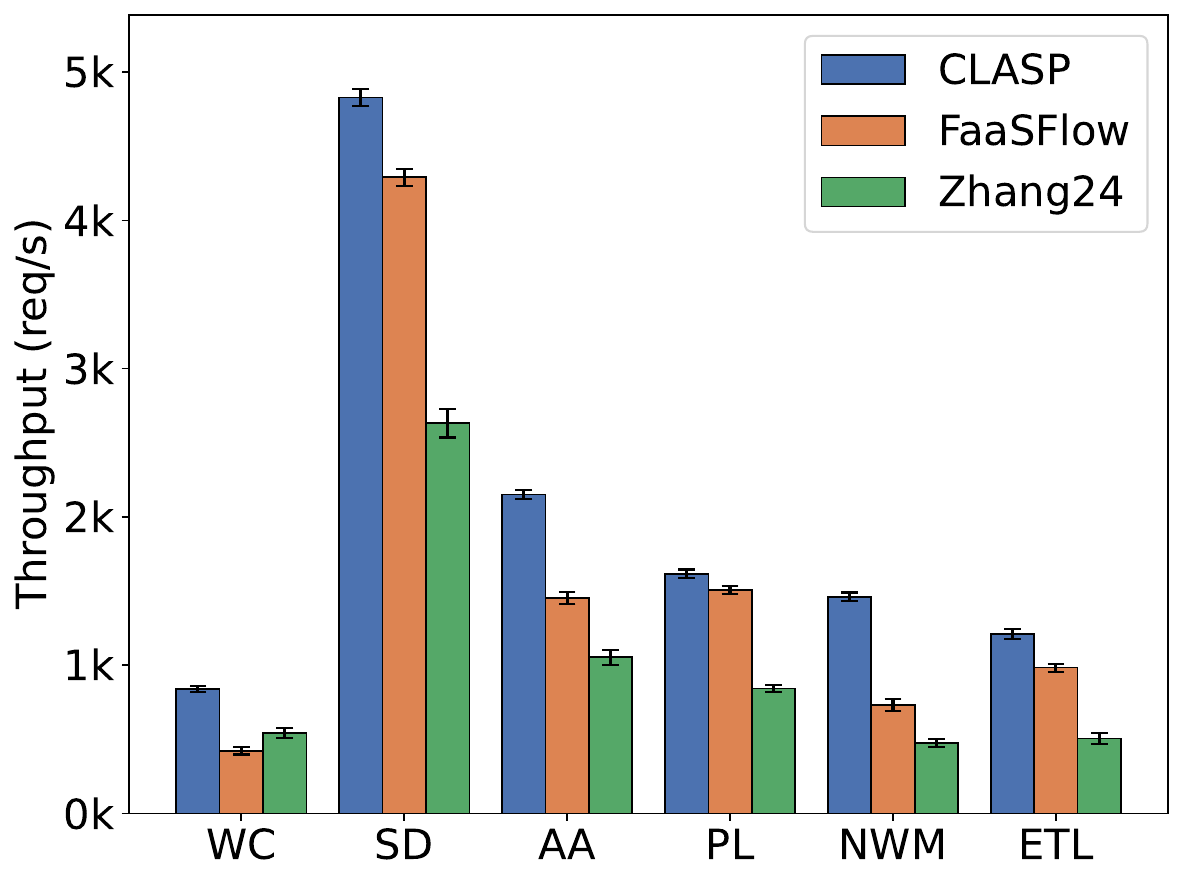}
\caption{Comparison of throughput across different applications.}
\Description{A grouped bar chart comparing three placement strategies on a fixed 10-worker cluster under unthrottled input. The horizontal axis lists six applications: WC, SD, AA, PL, NWM, and ETL. The vertical axis is throughput in requests per second, from 0 to about 5000. Each application has three bars: CLASP, FaaSFlow, and Zhang24. CLASP is the tallest bar in every application. The largest absolute gap is in SD, where CLASP reaches nearly 5000 against about 4200 for FaaSFlow and about 2600 for Zhang24. WC shows the largest relative drop for FaaSFlow, whose bar falls to roughly half of CLASP's. Zhang24 is the lowest bar in every application except WC.}
\label{fig:4.3.throughput_10}
\end{figure}

In this section, we evaluate our operator placement strategy Section~(\ref{sec:reverse_placement}) against the placement strategies of FaaSFlow and Zhang24. We run each application with unthrottled input on 10 workers, and compare their throughput for different methods.

As shown in Figure~\ref{fig:4.3.throughput_10}, our method achieves higher throughput than both baselines across all applications when running on ten workers. 

FaaSFlow falls behind because it does not account for the cost of chained requests. Although it reduces the number of remote chained requests by co-locating heavily communicating operators on the same worker, it does not quantify the chained-request cost that each operator imposes when placed onto a worker. As a result, this unaccounted cost leaves the workload imbalanced across workers and reduces throughput. This effect is most pronounced in WC, whose first operator emits more than ten chained requests for each request it processes. Since FaaSFlow does not model the cost of chained requests, it packs more instances of this operator onto a worker than the worker can actually sustain. The worker hosting these instances becomes overloaded, and WC suffers the largest throughput drop.

Zhang24 performs worse for a different reason: it spreads each operator's instances across workers to achieve workload balance. As a result, the instances of an operator's successor are scattered over many workers, so a single worker may have to send data to nearly every other worker, incurring a heavy fan-out cost.

\subsection{Worker Number Prediction}
\label{sec:exp_host_prediction}

\begin{figure*}[t]
\centering
\begin{subfigure}[b]{0.32\textwidth}
    \centering
    \includegraphics[width=\textwidth]{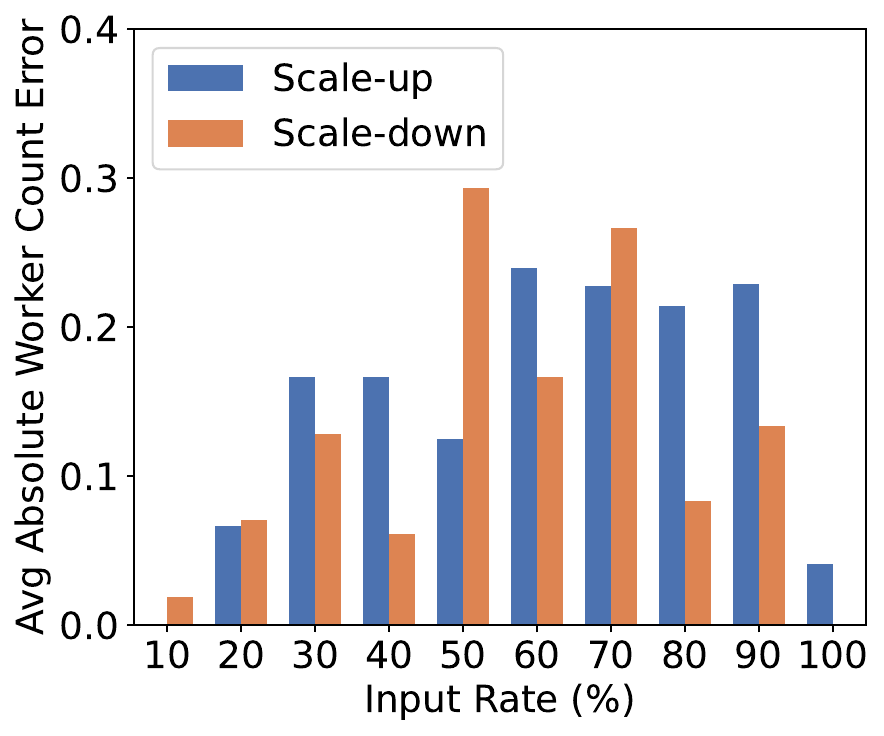}
    \caption{Comparison of absolute worker count error across different target rate.}
    \label{fig:4.3.our_scale_error_input}
\end{subfigure}
\begin{subfigure}[b]{0.32\textwidth}
    \centering
    \includegraphics[width=\textwidth]{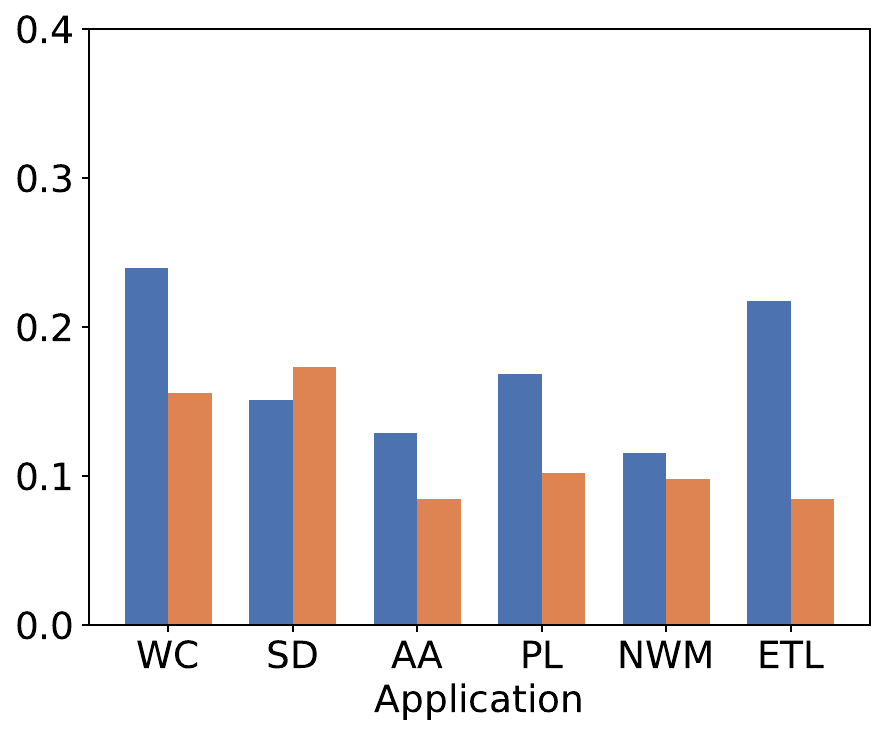}
    \caption{Comparison of absolute worker count error across different applications.}
    \label{fig:4.3.our_scale_error_bar}
\end{subfigure}
\begin{subfigure}[b]{0.32\textwidth}
    \centering
    \includegraphics[width=\textwidth]{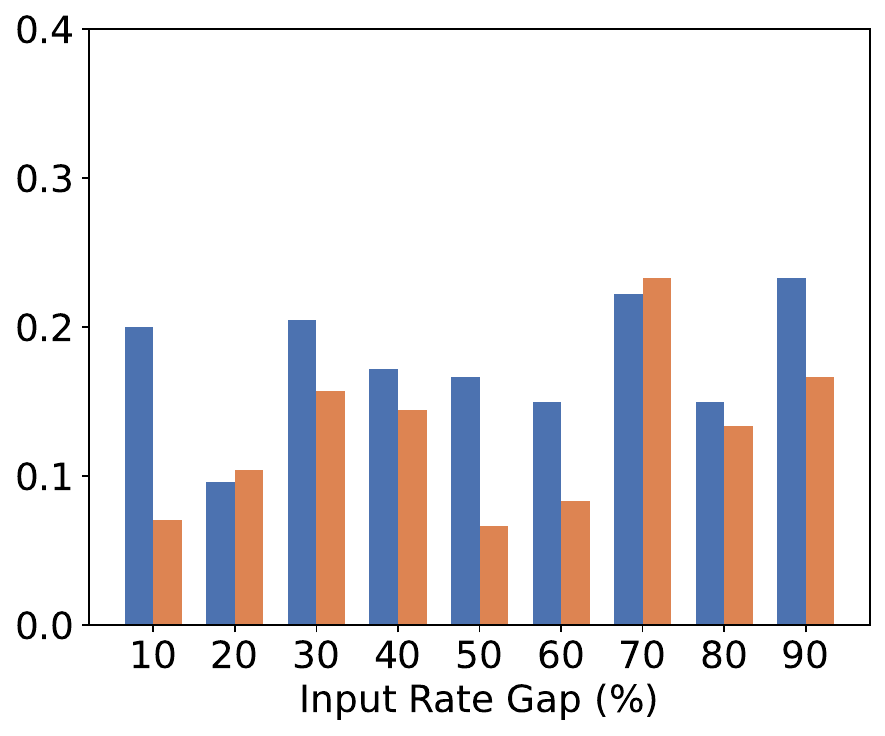}
    \caption{Comparison of absolute worker count error across different input rate gap.}
    \label{fig:4.3.our_scale_error}
\end{subfigure}
\caption{CLASP worker-count prediction error.}
\Description{Three bar charts side by side, all sharing a vertical axis of average absolute worker-count error from 0 to 0.4, and all distinguishing scale-up from scale-down with paired bars. Panel (a) plots error against target input rate, from 10 percent to 100 percent. Most bars sit below 0.2; the scale-up bars peak near 0.3 at a 50 percent target rate, and scale-down peaks near 0.27 at 60 percent. Panel (b) plots error per application for WC, SD, AA, PL, NWM, and ETL. All bars stay below 0.25, with WC and ETL highest on scale-up and AA lowest overall. Panel (c) plots error against the gap between source and target input rate, from 10 percent to 90 percent. All bars stay below 0.3 and show no upward trend as the gap widens. Across all three panels the scale-down bars are on average slightly shorter than the scale-up bars.}

\end{figure*}

\begin{figure}[t]
\centering
\includegraphics[width=0.65\textwidth]{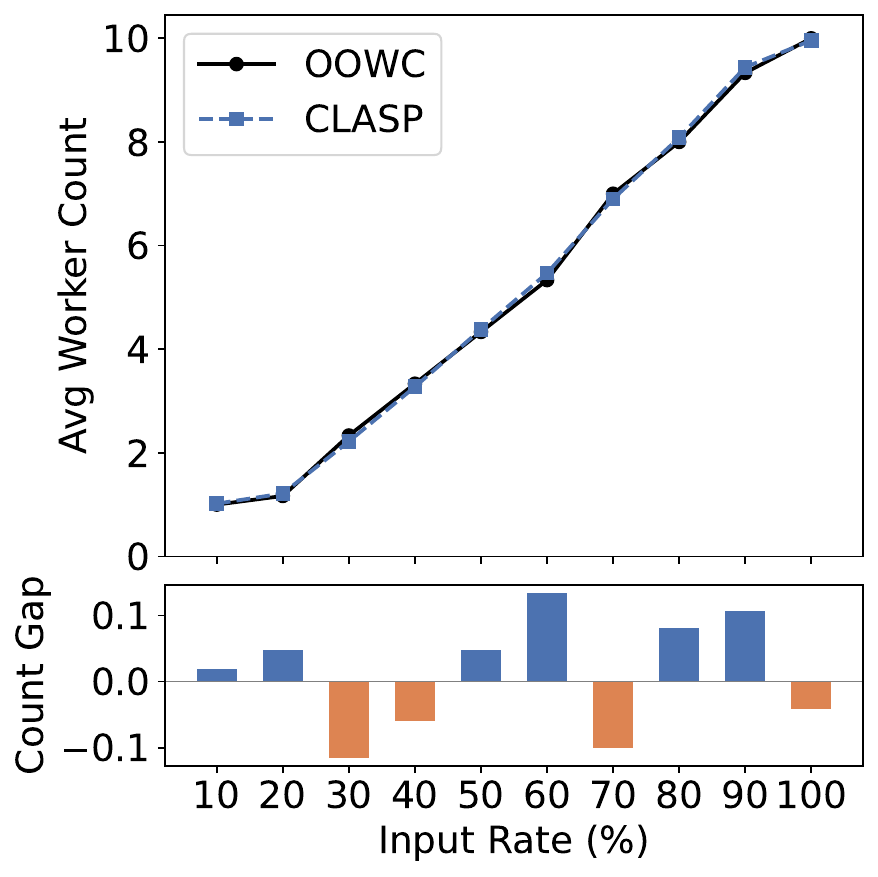}
\caption{Comparison of worker counts chosen by CLASP versus the observed optimal worker count (OOWC), where the count gap is defined as CLASP - OOWC.}
\Description{Two stacked panels sharing a horizontal axis of target input rate from 10 percent to 100 percent. The upper panel plots average worker count from 0 to 10 as two nearly coincident lines, one for the observed optimal worker count and one for CLASP. Both rise monotonically from about 1 worker at a 10 percent input rate to 10 workers at 100 percent, and the two lines are visually indistinguishable at most points. The lower panel plots the count gap, CLASP minus the observed optimal, as a bar chart on a vertical axis spanning roughly minus 0.1 to plus 0.15. The bars alternate in sign and never exceed about 0.13 in magnitude, showing that CLASP tracks the optimum closely and errs slightly high at some rates and slightly low at others.}
\label{fig:4.3.our_worker_deviation}
\end{figure}

In this experiment, we evaluate the accuracy of the worker count predicted by our scaling policy in Section~(\ref{sec:worker_scaling_policy}). After getting the maximum throughput for each application using our method on ten workers, we discretize this maximum into ten input-rate levels, 10\%, 20\%, ..., 100\%. For each application, we measure the maximum throughput sustained under each worker count from 1 to 10. For each input-rate level, we then take the smallest worker count whose maximum throughput meets or exceeds that level as the \emph{observed optimal worker count} (OOWC). We then evaluate every pairwise transition among the ten levels: starting from each source rate, we switch to each of the other nine levels as the target rate, and compare the predicted worker count against the observed optimal worker count at the target rate.

Figure~\ref{fig:4.3.our_worker_deviation} shows the average number of workers scaled under different target input rates, averaged over different applications and source rate. As shown, our method closely tracks the observed optimal worker count with only minor deviation across all target input rates. Figure~\ref{fig:4.3.our_scale_error_input} shows the absolute worker count error under different target rates; in most cases, the scaling worker error is less than 0.2. These two figures show, on average, our method follows the same pattern as the observed optimal worker count, only fluctuating slightly around it.

Figure~\ref{fig:4.3.our_scale_error_bar} reports the absolute worker count error relative to the observed optimal worker count, for different applications. Across all applications, the absolute worker count error stays below 0.25 workers, indicating that our method performs reliably.

Figure~\ref{fig:4.3.our_scale_error} reports the absolute worker count error relative to the observed optimal under different input-rate gaps, averaged over different applications and source/target rates. Here the gap is the difference between the source and target rates; since the rates span 10\% to 100\%, the largest gap is 90\%. Our method shows consistently similar error across all gaps, staying below 0.3 workers throughout. This indicates that its accuracy remains stable even under large, abrupt changes in the input rate. Our method scales more accurately when scaling down than when scaling up, though the difference is small.

\subsection{End-to-End Performance}

\begin{figure}[t]
\centering
\includegraphics[width=0.65\textwidth]{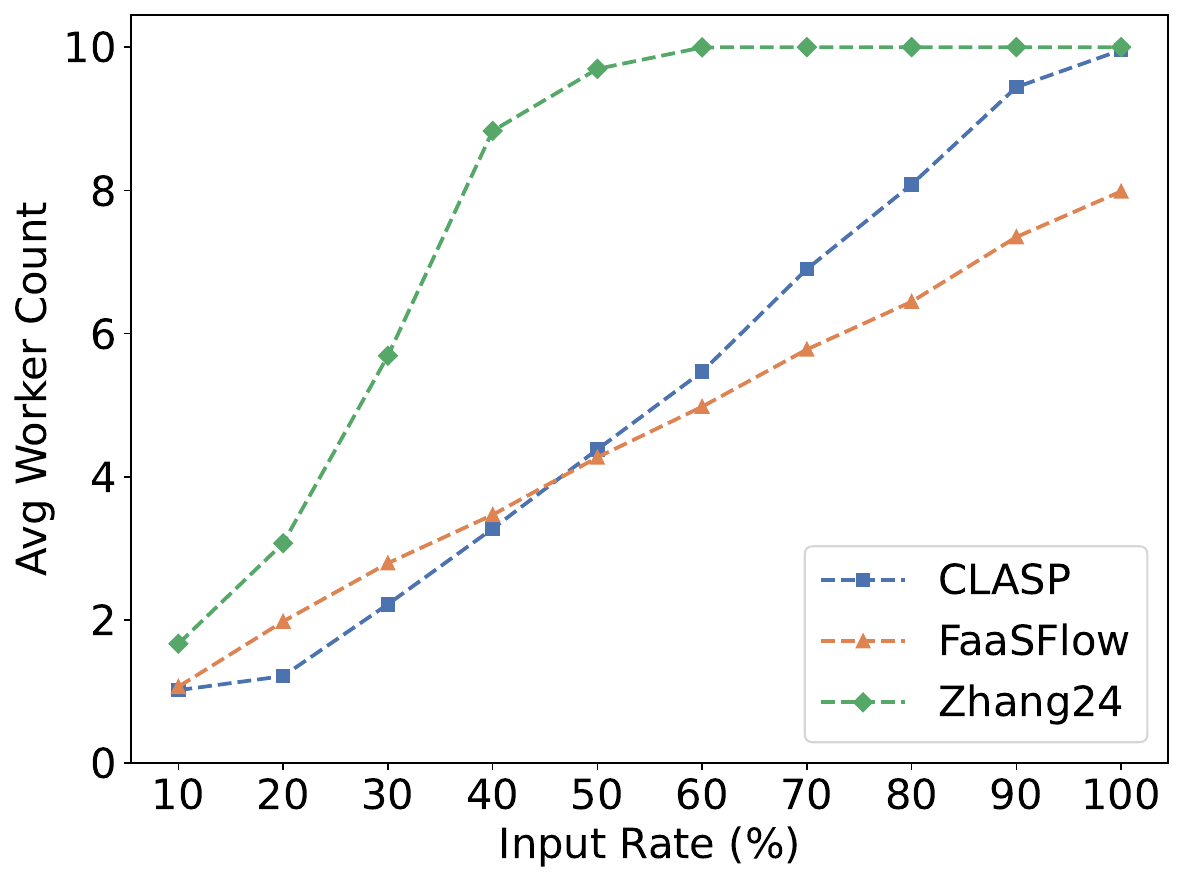}
\caption{Comparison of predicted worker counts under different input rates.}
\Description{A line chart with three series: CLASP, Zhang24, and FaaSFlow. The horizontal axis is input rate from 10 percent to 100 percent; the vertical axis is average worker count from 0 to 10. Zhang24 rises steeply, reaching nearly 9 workers at a 40 percent input rate and flattening at the 10-worker ceiling from 50 percent onward, so it over-provisions across most of the range. CLASP rises roughly linearly from about 1 worker at 10 percent to 10 workers at 100 percent. FaaSFlow tracks CLASP closely up to about 40 percent, then falls below it, ending at about 8 workers at a 100 percent input rate and so under-provisioning at high rates.}
\label{fig:4.5.scaling_trending}
\end{figure}

\begin{figure*}[t]
\centering
\includegraphics[width=\textwidth]{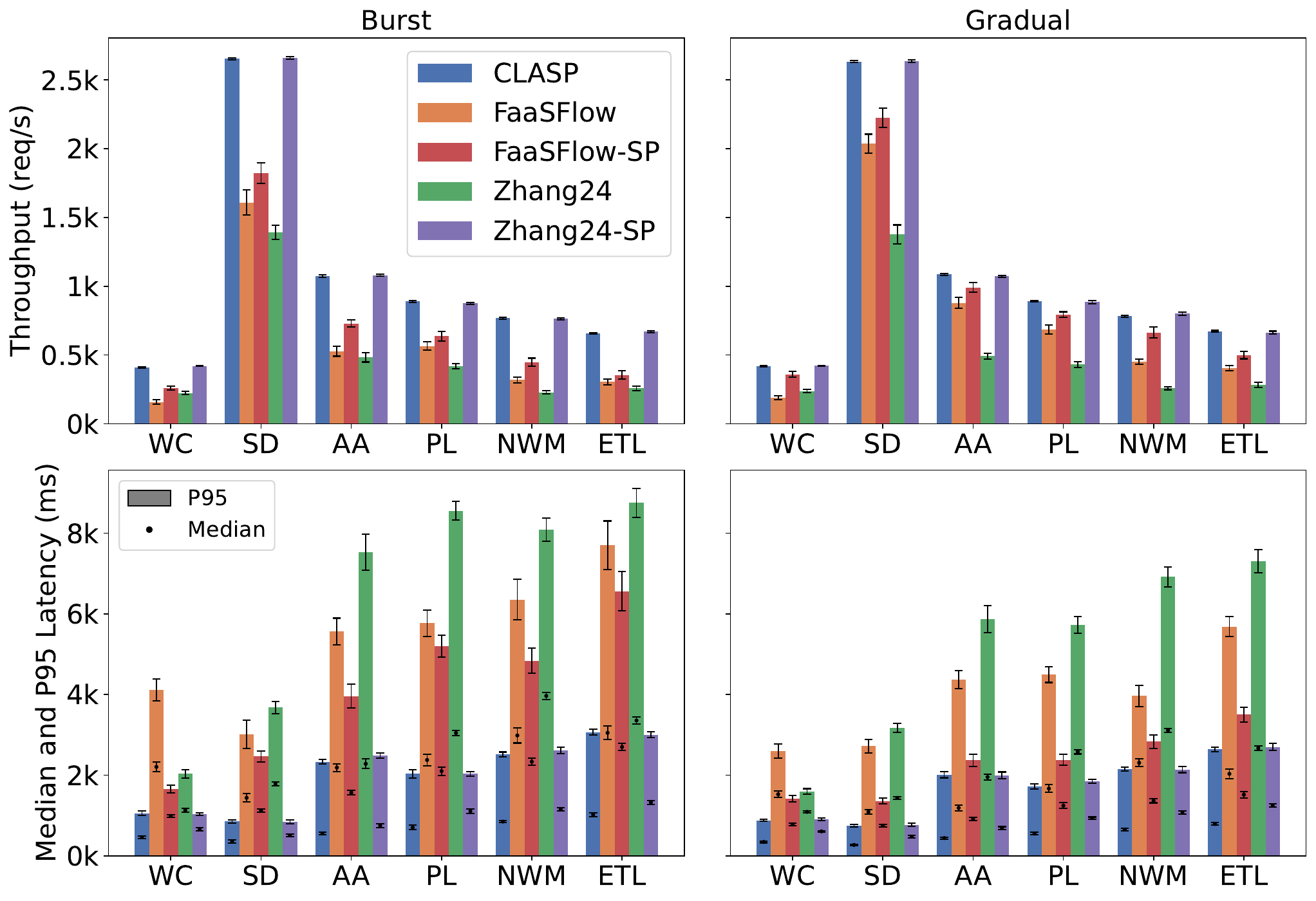}
\caption{Latency and Throughput Performance across different applications}
\label{fig:4.4.method_throughput_latency}
\Description{A two-by-two grid of grouped bar charts. The left column is the Burst input mode and the right column the Gradual mode. The top row plots throughput in requests per second, from 0 to about 2600; the bottom row plots median and p95 latency in milliseconds, from 0 to about 8000, with each bar spanning to the p95 value and a diamond marker showing the median. All four panels share a horizontal axis of six applications: WC, SD, AA, PL, NWM, and ETL. Each application has five bars: CLASP, FaaSFlow, FaaSFlow-SP, Zhang24, and Zhang24-SP. In the throughput panels, CLASP and Zhang24-SP are the tallest and are close to each other, while FaaSFlow and Zhang24 are consistently lower; FaaSFlow-SP sits between them and is shorter in Burst mode than in Gradual mode. In the latency panels, CLASP has the shortest bars throughout, with Zhang24 the tallest, reaching about 8500 milliseconds for AA and ETL in Burst mode. Zhang24-SP's median markers sit consistently above CLASP's while its p95 values are close to CLASP's.}
\end{figure*}

In this experiment, we examine how the accuracy of the predicted worker count and the operator placement affect application performance. We run each application in two modes, Burst and Gradual. In Burst mode, the input rate jumps randomly between a low range (10\%--30\%) and a high range (70\%--100\%), with each stage lasting 5 to 10 seconds. In Gradual mode, the input rate first rises step by step from 10\% to 100\% in increments of 10\%, then falls back from 100\% to 10\% in decrements of 10\%, with each stage lasting 5 to 10 seconds. Although the input rates and stage durations are random, we fix the random seed so that all baselines see the same input-rate pattern. We also ensure that the total input volume over the experiment period is the same for both Burst and Gradual modes.

We compare our approach against four baselines. The first two are FaaSFlow~\cite{li2022faasflow} and Zhang24~\cite{zhang2024efficient}. However, as discussed in Section~(\ref{sec:exp_operator_placement}), even with the same number of workers, these baselines differ in throughput because of differences in operator placement. To isolate this factor, we add two more baselines, FaaSFlow-SP and Zhang24-SP: each uses its original scaling method to decide the number of workers, but then adopts the same operator placement as CLASP. This lets us ablate the effect of operator placement. 

We also run the same worker-count prediction experiment from Section~(\ref{sec:exp_host_prediction}) on FaaSFlow and Zhang24 to examine how each predicts the worker count under different input rates. As shown in Figure~\ref{fig:4.5.scaling_trending}, the three methods exhibit clearly different scaling trends.

Zhang24 tends to over-provision: it attempts to use all available workers even when processing only 50\% of the maximum rate, and its predicted worker count stops growing beyond 60\% because it has already reached the maximum number of workers. FaaSFlow's prediction curve is more linear, because it decides the worker count based solely on the number of processed requests, without accounting for the chained-request cost. Compared with our method, it uses more workers at low input rates but fewer at high input rates.

A key characteristic of FaaSFlow is that its worker-count prediction depends heavily on the source rate: for the same target rate, different source rates can lead to very different predictions. This is because FaaSFlow estimates the required worker count by linearly extrapolating from the currently observed throughput. However, even if FaaSFlow under-provisions in one scaling round, it monitors the current metrics and adjusts the worker count in the next round, so it does not stay under-provisioned for the same target rate. As a simplified example, consider an application running on a single worker with a throughput of 100. If the input rate rises to 500, FaaSFlow predicts that 5 workers are needed. After scaling to 5 workers, it observes a throughput of only 250, and therefore revises its prediction to 10 workers for the same target rate. Its prediction is thus poor when the source rate is far from the target rate, and improves as the source rate moves closer to it. Note that the curve for FaaSFlow in Figure~\ref{fig:4.5.scaling_trending} averages over both scaling directions, which hides a systematic bias: when scaling up, FaaSFlow consistently uses fewer workers than OOWC, and when scaling down, it consistently uses more. Neither CLASP nor Zhang24 shows this asymmetry between scaling up and scaling down.

As shown in Figure~\ref{fig:4.4.method_throughput_latency}, in both burst and gradual mode, our method consistently achieves higher throughput and lower latency than FaaSFlow and Zhang24. Our method achieves up to $2.5\times$ the throughput of FaaSFlow and $3.3\times$ that of Zhang24, and reduces median latency by up to 75\% relative to FaaSFlow and 76\% relative to Zhang24. This improvement comes from both the scaling method and the operator placement.

We also compare against FaaSFlow-SP and Zhang24-SP, which share the same placement as CLASP but retain their original scaling methods. Compared with FaaSFlow-SP, our method still consistently achieves better performance, because FaaSFlow-SP tends to under-provision resources at high input rates, which lowers throughput and increases latency. Moreover, FaaSFlow-SP's throughput is lower in burst mode than in gradual mode, because its prediction accuracy degrades when the source rate is far from the target rate. CLASP does not suffer from this, which shows that our method is robust in both burst and gradual modes. Compared with Zhang24-SP, our method achieves similar throughput across benchmarks, and the error bars are small and comparable for both, because both always provision enough resources to match the input rate and process all incoming requests. However, our method sustains this throughput with fewer workers. Both latencies are computed over all requests in a run. The input rate varies across levels during a run, so the pooled distribution mixes requests from every level. Latency is highest at the top input rate. The p95 is therefore dominated by requests issued at that level. There, both methods use all ten workers under the same placement, so their p95 latencies are close. The median behaves differently. It also reflects the lower and intermediate rates. At those rates, Zhang24-SP over-provisions, and the extra workers add inter-worker communication cost.

\section{Conclusions and Future Work}

We proposed CLASP, an elastic scaling and scheduling framework tailored for stream processing applications in stateful serverless environments. CLASP builds a worker capacity model that jointly captures execution cost and chained-request costs, and uses this model to schedule operators onto workers so as to minimize the number of workers used while sustaining the input rate. In addition, CLASP provides a migration strategy that migrates operator state and pending requests according to the new placement, minimizing the processing pause caused by rescheduling. Experimental results show that CLASP improves throughput and reduces end-to-end latency compared with state-of-the-art scaling strategies under both burst and gradual input-rate changes.

In future work, we plan to extend CLASP to heterogeneous clusters and to explore SLO-aware scaling objectives beyond minimizing the number of workers.

\textbf{Software Availability:}
The CLASP implementation will be released as open source after the paper is accepted.

\bibliographystyle{ACM-Reference-Format}
\bibliography{reference_papers}

\end{document}